\documentstyle[aps,prd,epsfig]{revtex}
\begin{document}
\draft
\title{Self-consistent treatment of bubble nucleation\\
at the electroweak phase transition}
\author{Arnd S\"urig
\footnote{e-mail:~suerig@hal1.physik.uni-dortmund.de}
\footnote{Supported by Deutsche Forschungsgemeinschaft (DFG) under
contract: Ba 703/4-1}}
\address{Institut f\"ur Physik, Universit\"at Dortmund\\
         D - 44221 Dortmund , Germany}
\date{\today}
\maketitle
\begin{abstract}
In the standard procedure for calculating the decay rate of a metastable
vacuum the solution of the classical Euclidean equation of motion of the
background field is needed. On the other hand radiative corrections have
to be taken into account already in the equation of motion.
Hence, the latter one has to be the functional derivative of the
effective action with respect to the background field. This is of 
crucial importance in theories in which the symmetry breaking is
due to radiative corrections. Usually the effective potential is 
considered only, neglecting the corrections due to the derivative 
terms of the effective action. In this article a bounce solution
from an equation of motion which takes into account the full effective
action in the one-loop approximation is calculated. 
A computational method that yields
a strict separation of the divergent contributions to the effective action
from the convergent ones is obtained. This allows a wide freedom in
the choice of regularization and renormalization schemes.
The model under consideration is the SU(2)-Higgs model. The fluctuations
of the complete bosonic sector, i.~e. gauge field, Higgs and Goldstone boson
contributions, are taken into account. The bounce is then self-consistent 
to one-loop order. The obtained results for characteristic quantities of the
transition as the nucleation rate and the number of nucleated bubbles 
per volume are compared to other, non-self-consistent approaches.
\end{abstract}
\pacs{11.15.Kc, 11.10.Gh, 11.10.Wx, 98.80.Cq}
\section{Introduction}
\label{einleitung}
When dealing with field theory in cosmological context, one often
encounters the problem that a field is not in a state which is the
absolute minimum of the potential, the true vacuum, but in a metastable
local minimum, the false vacuum, which is at higher free energy. The
transition from false to true vacuum proceeds by
a quantum mechanical tunnelling
process or by a classical transition induced
by thermal fluctuations for a system at finite temperature. The transition
is local in space. During the transition
a bubble is formed which is a region in space with true vacuum
surrounded by false vacuum.
The bubble just large enough not to collapse
is called critical bubble. Once nucleated, the bubble expands,
converting false vacuum to true vacuum.

If the mass of the Higgs boson is not too large the electroweak phase
transition is of first order and proceeds by bubble nucleation. This
scenario is interesting because of the possibility to explain the
baryon asymmetry within the minimal standard model
\cite{Kuzmin:1985,Cohen:1993}.

A quantity of fundamental importance for a phase transition
is the transition rate. The nucleation rate per volume, $\gamma$, can
be calculated from the bounce-solution of the Euclidean field equation.
This solution is also called classical solution. For a theory with only one
scalar field $\Phi$ with a potential $U(\Phi)$ one has to solve
the equation of motion:
\begin{equation}
\label{treelevequat}
\partial_\mu \partial_\mu \Phi - \frac{{\rm d}}{{\rm d} \Phi} U(\Phi) = 0\;.
\end{equation}
The boundary condition is, that the solution $\Phi=\phi$ tends to its
false vacuum value if one of the $x_\mu$ tends to $\infty$. The nucleation
rate can be written in the form \cite{Coleman:1977,Callan:1977}
\begin{equation}
\gamma= A \exp(-B)\;,
\end{equation}
where $B$ is the Euclidean action of the classical solution $\phi$:
\begin{equation}
\label{sclbounce}
B = \int \! {\rm d}^4x \, \left \{ \frac{1}{2} (\partial_\mu \phi)^2
+ U(\phi) \right \}\;.
\end{equation}
In the semiclassical approximation the factor $A$ is given
by an expression involving functional determinants 
\cite{Langer:1967,Langer:1969,Affleck:1981}.
The negative
logarithm is the one-loop contribution to the effective
action.

If $U(\Phi)$ only has a single minimum the bounce-solution does not
exist. Nevertheless, in some models
the vacuum structure is not determined by $U(\Phi)$ alone but it
is changed by radiative corrections. These can be calculated by
taking into account the quantum fluctuations of the scalar field
and those fields coupled to the scalar one \cite{Coleman:1973}.
Depending on the specific model, the radiative corrections yield
an effective potential $V_{\rm eff}(\Phi)$ with different minima
\cite{Dine:1992,Kirzhnits:1976,Shaposhnikov:1987,Shaposhnikov:1988,Anderson:1992,Kripfganz:1995,Kripfganz:1997}.
The bounce-solution is then calculated from a field equation
that contains this effective potential.
However, the effective potential
is only a part of the effective action. It does not take into account
the space-time dependence of the background fields, i.~e.\ terms, that
depend on derivatives of the background fields.

This raises the question, whether it is justified to neglect these terms
in calculating the bounce \cite{Weinberg:1993}. 
In case of the electroweak phase transition
the bosonic as well as the fermionic contributions to the effective action
have been calculated, based on a bounce-solution, that was determined
from the effective potential and not the effective action 
\cite{Baacke:1995,Baacke:1996}. In
\cite{Kripfganz:1995} an improvement to take into account non-local effects
already for the calculation of the bounce was done by modifying
the kinetic term of the background field. The radiative corrections
were calculated using approximations as the derivative expansion
\cite{Baacke:1994} and the heat kernel method \cite{Kripfganz:1995} 
or by exact numerical computation schemes
\cite{Baacke:1995,Baacke:1996,Baacke:1994,Baacke:1993}. The corrections 
due to the bosonic sector (gauge and Higgs fields) were
found to be of the same order as the classical action. In such a case the
non-local terms of the effective action are not negligible and should be
taken into account when calculating the classical solution. As the
fermionic corrections are small compared to the classical 
action \cite{Baacke:1996}, their
non-local contributions to the effective action
can be omitted in the equation of motion.

The standard procedure to calculate the nucleation rate works in the
following way:
First one adds the one-loop corrections of the effective potential
to the classical action $B$ and gets an approximation 
$\widetilde S$ of the one-loop effective action. Based upon the
functional form of $\widetilde S$ the field equation is derived and
the bounce is determined. This is used to evaluate $\widetilde S$ and the
prefactor $A$ of the transition rate which is given in more detail by
\begin{equation}
  \label{nuclrate}
  \gamma=\frac{\omega_-}{2\pi} \left( \frac{\widetilde S}{2\pi} 
  \right)^{\frac{3}{2}}\exp\left(- \widetilde S - \frac{1}{2}
    \ln {\cal J}_B \right)\;.
\end{equation}
Here $\omega_-$ is the modulus of the eigenvalue of the bubble's unstable mode.
${\cal J}_B$ denotes the product of all fluctuation determinants of
the bosonic sector. Aside the dominant fermionic contribution due to the
top-quark which
is usually taken into account for the setup of $\widetilde S$
the fluctuation determinant due to fermions is neglected here.
The aim of this paper is to improve the calculation of the bounce
which is then used as the numerical input for the evaluation of
the nucleation rate. It is no longer determined as a saddle point
of $\widetilde S$ but as a saddle point of the full one-loop effective
action. I.~e., it will be obtained as a self-consistent solution of
the one-loop field equation:
\begin{equation}
  \label{olfeq}
  \frac{\delta}{\delta \phi(\mbox{\boldmath$x$\unboldmath})}
  \Gamma^{{\rm 1}\mbox{-}{\rm l}}=
  \frac{\delta}{\delta \phi(\mbox{\boldmath$x$\unboldmath})} 
  \left( \widetilde S 
    + \frac{1}{2}\ln {\cal J}_B \right) =0\;.
\end{equation}

The problem is, that no analytic representation of the
effective action in the one-loop approximation is known.
In \cite{Baacke:1997}
a numerical method to calculate the functional derivative of an effective
action has been presented. It is based upon a computation scheme to
determine full one-loop Green's functions 
\cite{Baacke:1990a,Baacke:1990b,Baacke:1992}. In this paper
this technique is applied to determine a bounce-solution from a field equation
where the full space-time-dependence of the non-local terms
of the quantum corrections to one-loop
order is taken into account. As the numerical method to compute the functional
derivative needs the background field as a numerically known input quantity
the procedure we use to obtain the background field is iterative. It finally
results in a self-consistent fixed point of the iteration condition, i.~e. a
solution of the one-loop field equation (\ref{olfeq}).

A one-loop Green's function is a divergent quantity that
has to be renormalized. An important issue of the presented method is
the strict separation of the divergent parts, which have to be handled
analytically, and the convergent part, which may be computed numerically.
As no expansion in any quantity is needed a 
conceptual advantage of this method in contrast to
approximation schemes like the derivative or the heat kernel expansion
is that the unphysical infrared divergencies these approximations suffer from
do not appear.

Among the various models for which
self-consistent solutions are interesting this paper deals with
the electroweak phase transition. Recent lattice calculations show
that the perturbation theory is not reliable for Higgs boson masses
above $60$GeV. 
The phase transition seems to disappear at a critical
Higgs boson mass which is in the range between $66$GeV and $80$GeV
\cite{Gurtler:1997hr} and the experimental lower limit of the Higgs
boson mass already is beyond $66$GeV.
Hence, the generation of the baryon asymmetry at the phase
transition of the standard model is ruled out already, and the
interest has shifted to extensions of the standard model.
Nevertheless, the investigation of the first order phase transition
at small Higgs boson masses is still of interest and is at present
actively studied on the lattice \cite{Moore:1997bn}.
Some aspects of the transition, like
the determination of the surface tension and the latent heat, are
still in a state of development. Hence, we have the
possibility to develop new calculational methods in a well known
model that are straightforward to extend to more complicated
models.

The plan of this paper is as follows: In the next section the model will
be specified and the relevant fluctuation operators are given.
The next section deals with
the renormalization of the effective action that is improved by
resummation.
In section~\ref{sec:numericmethods} the formulas for the
numerical procedure how to calculate the functional derivative of an effective
action and the effective action itself are derived.
Finally the results are presented and discussed.
%
%
\section{Basic Equations}
\label{sec:basicequations}
The Lagrangian of the SU(2)-Higgs model with gauge fields and fermions
in the limit of vanishing electroweak mixing angle is given by
\begin{equation}
  \label{lstandmod}
  {\cal L} = {\cal L}_{\rm gauge} + {\cal L}_{\rm Higgs} 
  + {\cal L}_{\rm ferm}\;,
\end{equation}
where
\begin{eqnarray}
  {\cal L}_{\rm gauge}&=&-\frac{1}{4} W_{\mu\nu}^a W^{\mu\nu a}\; , \\
  {\cal L}_{\rm Higgs}&=& \frac{1}{2} ({\rm
    D}_\mu \Phi) ^\dagger ({\rm D}^\mu \Phi) - U(\Phi^\dagger\Phi)\; ,\\
  {\cal L}_{\rm ferm}&=& \sum_f \bar \Psi^f i \gamma^\mu \partial_\mu
  \Psi^f - \sum_{ff^\prime} g_Y^{ff^\prime} \Phi \bar \Psi^f
  \Psi^{f^\prime}\;,\\
  W_{\mu\nu}^a &=& \partial_\mu A_\nu^a - \partial_\nu A_\mu^a +
  g\varepsilon^{abc}A_\mu^b A_\nu^c \; ,\\
  {\rm D}_\mu &=& \partial_\mu - i\frac{g}{2} \tau^a A_\mu^a\; ,\\
  U(\Phi^\dagger \Phi) &=& \frac{\lambda}{4}\left( \Phi^\dagger\Phi - v_0^2
  \right)^2 \;.
\end{eqnarray}
Here $W_{\mu\nu}^a$ denotes the field strength tensor of the gauge fields,
$\Phi$ the Higgs-doublet, $\tau^a$ the Pauli matrices, $\Psi^f$ 
the fermion fields of the quarks,
$v_0$ the vacuum expectation value of the Higgs field at temperature
$T=0$ and $g_Y^{ff^\prime}$ the Yukawa-couplings of the fermions to the
Higgs field. The sum $f$ runs over the different
flavors and colors of the quarks.

The critical electroweak bubble is a pure real Higgs field
configuration. On the classical level it is
\begin{equation}
\label{classfieldconfig}
\Phi(x)_{\rm cl} = \phi(x)\left( 0 \atop 1 \right)\;, \qquad
A_\mu^a(x)_{\rm cl}=0\;, \qquad \Psi(x)_{\rm cl}=0\;.
\end{equation}
The configuration is taken to be stationary, i.~e.\ 
time independent. As explained
in the introduction the bubble profile calculated from the functional
derivative of the classical action (\ref{sclbounce}) does not exist
because the tree-level potential
$U(\phi)$ only has a single minimum. Hence, instead
of the classical action one has to use the effective action. For the model
under consideration this effective action is given by:
\begin{equation}
  \Gamma[\phi,T]=S_{\rm cl}[\phi] + S_{\rm eff,gauge}[\phi,T] +
  S_{\rm eff,ferm}[\phi,T]+ S_{\rm eff,Higgs}[\phi,T]\;.
\end{equation}
This functional cannot be computed exactly. Usually it is calculated in
the semiclassical approximation to one-loop order.
In order to calculate the one-loop approximation
one needs to know the fluctuation operators. They are found by
expanding the Lagrangian (\ref{lstandmod}) 
around the classical field configuration (\ref{classfieldconfig})
to second order in the small fluctuations:
\begin{eqnarray}
  \Phi(x) &=& \left[ \phi(x) + h(x) + i \tau^a\varphi^a(x) \right]
  \left( 0 \atop 1\right)\; ,\nonumber\\
  A_\mu^a(x) &=& a_\mu^a(x)\;,\\
  \Psi(x)&=& \psi(x)\nonumber\;.
\end{eqnarray}
Here $h(x)$, $\varphi^a(x)$, $a_\mu^a(x)$ and $\psi(x)$
denote the fluctuations of the
isoscalar-- and isovector--part of the Higgs field, the gauge field
and the fermions respectively. The second order contribution to the
Lagrangian is given by:
\begin{eqnarray}
  {\cal L}^{(2)} &=& -\frac{1}{2}\partial_\mu a_\nu^a \partial^\mu a^{\nu a}
  +\frac{1}{2} \partial_\mu a_\nu^a \partial^\nu a^{\mu a}
  + \frac{g^2}{8} \phi^2a_\mu^a a^{\mu a}
  +\sum_f \bar \psi^f i \gamma^\mu \partial_\mu
  \psi^f - \sum_{ff^\prime} g_Y^{ff^\prime} \phi \bar \psi^f
  \psi^{f^\prime}\nonumber\\
  && + \frac{1}{2}\partial_\mu h \partial^\mu h 
  + \frac{1}{2}\partial_\mu \varphi^a
  \partial^\mu \varphi^a + \frac{g}{2}(\partial_\mu \phi) a^{\mu a} \varphi^a
  -\frac{g}{2}\phi a_\mu^a\partial^\mu \varphi^a\\
  &&-\frac{1}{2}\left( 3 \lambda \phi^2 -\lambda v_0^2\right) h^2
  - \frac{1}{2} \left( \lambda \phi^2-\lambda v_0^2 \right) 
  \varphi^a \varphi^a\; ,\nonumber
\end{eqnarray}
One has to add a gauge-fixing term and the contributions of the
Faddeev-Popov-ghosts to the Lagrangian. Here the `t Hooft-Feynman-background
gauge is used:
\begin{eqnarray}
  {\cal L}_{\rm GF} &:=& -\frac{1}{2} {\cal F}^a {\cal F}^a \qquad
  \mbox{with} \qquad{\cal F}^a := \partial_\mu a^{\mu a} 
  + \frac{g}{2}\phi \varphi^a\;,\\
  {\cal L}_{\rm FP} &=& \eta^{\dagger a}\left(\partial_\mu \partial^\mu 
    + \frac{g^2}{4}\phi^2 \right) \eta^a\;.
\end{eqnarray}
From
${\cal L}_{\rm tot}^{(2)}=\left( {\cal L} + {\cal L}_{\rm GF} + {\cal
    L}_{\rm FP}\right)^{(2)}$ the fluctuation operators are read of.
They are given in Euclidean metric as the bounce solution we want to
obtain is a solution of an Euclidean field equation. There are three
contributions of the bosonic sector to the effective action. The first
is due to the isoscalar fluctuations of the Higgs field, the second is
due to the coupled channel of the gauge-fields and the
isovector components of the Higgs field (would-be-Goldstone bosons)
and the last is due to the time component of
the gauge fields and the Faddeev-Popov-ghosts. Each of them can be written as:
\begin{equation}
  S_{i}[\phi] = c_i \ln {\cal J}_i = c_i \ln \det \left( 
    \frac{-\partial^2 + {\cal U}_i(\phi)}{-\partial^2+{\cal U}_i(0)}
  \right)\;.
\end{equation}
In order to shorten the notation the quantities
\begin{eqnarray}
  \label{degen}
  c_{a_0 \eta}&=&-\frac{3}{2}\;, \qquad c_h=\frac{1}{2} \;, \qquad
  c_{a\varphi}= \frac{3}{2}\;,\nonumber\\
  \label{mh2phi}
  m_a^2(\phi)&=&\frac{g^2}{4}\phi^2\;, \quad
  m_h^2(\phi)=3\lambda\phi^2-\lambda v_0^2\;, \quad
  m_\varphi^2(\phi)=\frac{g^2}{4}\phi^2 + \lambda\left(\phi^2-v_0^2\right)\;,\\
  \label{ua0eta}
  {\cal U}_{a_0 \eta}(\phi) &=& m_a^2(\phi)\;,\quad
  {\cal U}_h(\phi) = m_h^2(\phi)\;,\\
  {\cal U}_{a \varphi}(\phi) &=& \left( 
    \begin{array}{llll}
      m_a^2(\phi) & 0 & 0 & 0\\
      0 & m_a^2(\phi) & 0 & 0\\
      0 & 0 & m_a^2(\phi) & 0\\
      0 & 0 & 0 & m_\varphi^2(\phi)\\
    \end{array}
  \right) + g \mbox{\boldmath$\zeta$\unboldmath} 
  \mbox{\boldmath$\nabla$\unboldmath} \phi\;,\nonumber\\
  \mbox{\boldmath$\zeta$\unboldmath} &:=& \left( 
    \zeta^1 \quad \zeta^2 \quad \zeta^3 \right)^T\;,\\
  \zeta^1&:=& \left(
    \begin{array}{llll}
      0&0&0&1\\0&0&0&0\\0&0&0&0\\1&0&0&0\\
    \end{array}
  \right)\;,
  \ 
  \zeta^2:= \left(
    \begin{array}{llll}
      0&0&0&0\\0&0&0&1\\0&0&0&0\\0&1&0&0\\
    \end{array}
  \right)\;,
  \ 
  \zeta^3:= \left(
    \begin{array}{llll}
      0&0&0&0\\0&0&0&0\\0&0&0&1\\0&0&1&0\\
    \end{array}
  \right)\;.
\end{eqnarray}
are introduced and one uses that the background field is
time independent.

The contribution of the top-quark is the dominant one in the fermionic sector.
Therefore, nothing but its contribution is taken into account. It is given by
\begin{equation}
\label{stformdef}
S_t[\phi] = -3 \ln \det \left( 
\frac{\gamma_\mu \partial_\mu - g_t \phi}
{\gamma_\mu \partial_\mu} \right)\;,
\end{equation}
where the factor three is due to the quark color and $g_t$ is 
the strength of the Yukawa-coupling of the top-quark. As explained
in the introduction only the leading temperature dependence of the
top-quark contribution, which is part of the finite temperature effective
potential, will be taken into account. Hence, this calculation is not
self-consistent with respect to the fermionic sector.

The first step is to extract the contributions of the different fluctuating
fields to the high temperature approximation of the one-loop effective
potential \cite{Dine:1992,Anderson:1992,Carrington:1992,Buchmuller:1994bq}.
Based upon the fluctuation operators given above the sum of the tree
level potential $U(\phi)$ and these one-loop corrections 
can be written as:
\begin{eqnarray}
  \label{v_ges}
  V_{\rm ht}(\phi,T)&=&
  \frac{1}{2}\check m_H^2 \phi^2 - ET \phi^3 + \frac{\lambda_T}{4} \phi^4\\
  \mbox{with } \check m_H^2&:=& 2D (T^2 - T_0^2)\\
  \label{const_d}
  D&:=&\frac{1}{8v_0^2}\left( 3m_W^2 + m_H^2 + 2m_t^2\right)\\
  B&:=&\frac{3}{64\pi^2 v_0^4} \left(2m_W^4
    +\left(\frac{m_H^2}{2}+m_W^2\right)m_H^2 
    + \frac{7}{8}m_H^4 - 4m_t^4 \right)\\
  T_0^2&:=&\frac{m_H^2 - 8v_0^2 B}{4D}\\
  E&:=&\frac{3g^3}{32\pi}\\
  \lambda_T&:=& \lambda-\frac{3}{16\pi^2 v_0^4}
  \left[2m_W^4 \ln\frac{m_W^2}{a_BT^2}
    +\left(\frac{m_H^2}{2} + m_W^2 \right)^2 \ln \frac{m_W^2}{a_BT^2}\right.\\
  && \left. + 
    \frac{3}{4}m_H^4 \ln \frac{m_H^2}{a_BT^2} - 4m_t^4 \ln \frac{m_t^2}{a_FT^2}
  \right]\;,\nonumber
\end{eqnarray}
where $\ln a_B=2 \ln 4\pi - 2 \gamma \simeq 3.91$, $\ln a_F=2 \ln \pi -
2\gamma \simeq 1.14$, $v_0=246$GeV, $m_W=\frac{1}{2}gv_0$ and $\Theta_W=0$.
The form of the constants $B$, $D$, $E$ and $\lambda_T$ differs
from those in \cite{Dine:1992}, 
because of the contributions of the Higgs boson and
would-be-Goldstone boson fluctuation taken into account here. The
renormalization has been performed in the minimum of the classical
potential at $T=0$.

The given representations of the effective action are quite formal.
They are divergent and have to be renormalized. Furthermore, they have
to be evaluated for a system at finite temperature. This is done
in the next section
%
%
\section{Renormalization of the resummed effective action}
\label{sec:renormierung}
The vacuum structure is changed due to the contributions of the different
fluctuating fields to the one-loop effective potential. The tree-level mass
and the tree-level self-coupling of the 
Higgs field are modified due to effects of 
finite temperature. They will be modified again by effects of higher order.
Dependent on the temperature these corrections are not negligible
and a pure one-loop calculation is not reliable 
\cite{Linde:1979px,Gross:1981br}.

In order to take into account at least the dominant contributions of the
higher order effects one has to consider the sum of one-loop-Daisy-graphs,
replacing the expressions due to the inserted
tadpole-graphs by their leading temperature
dependence which is proportional to $T^2$.
Also the vertices are replaced by those which take into account
the temperature dependence of the coupling.

In practice the summation of Daisy-graphs is performed by replacing
the tree-level-potential $U(\phi)$
in the Lagrangian by the high temperature potential $V_{\rm ht}$
given in the preceding section
\cite{Weinberg:1974hy,Pisarski:1989vd,Braaten:1990mz,Parwani:1992}.
Then the fluctuation operators and the contributions to the effective
action are determined.
The effective action based upon these fluctuation operators is improved
by resummation. It can be written as
\begin{equation}
  S_{\rm R}[\phi,T]=\sum_{i} S_{i,\rm R}[\phi,T]=\sum_{i} c_i 
  \sum_{n=-\infty}^{+\infty} \ln \det
  \left(
    \frac{-\Delta + \nu_n^2 + {\cal U}_i(\phi,T)}{-\Delta + \nu_n^2 +
      {\cal U}_i(0,T)}
  \right)\;,
\end{equation}
where ${\cal U}_i(\phi,T)$ is given by (\ref{ua0eta})ff.\  with
$m_h^2(\phi)$ and $m_\varphi^2(\phi)$ replaced by
\begin{equation}
  m_h^2(\phi,T)=3 \lambda_T \phi^2 - 6ET\phi + \check m_H^2\;, \quad
  m_\varphi^2(\phi,T)=\frac{g^2}{4}\phi^2 + \lambda_T \phi^2 - 3ET\phi
  + \check m_H^2\;.
\end{equation}
Expanding the effective action in a series of one-loop Feynman-graphs
it can be rewritten as
\begin{eqnarray}
  \label{sirphitexpansion}
  S_{i,\rm R}\!\!\!&=&\!\!\!\sum_{k=1}^{+\infty} \frac{(-1)^{k+1}c_i}{k}
  \!\!\sum_{n=-\infty}^{+\infty} \!\! {\rm Tr}
  \left \{ \left[ -\Delta + \nu_n^2
      +{\cal U}_i(0,T)\right]^{-1} \left[ {\cal U}_i(\phi,T) 
      - {\cal U}_i(0,T)\right]
  \right\}^k\\
  &=& \!\!\!\sum_{k=1}^{+\infty} S_{i,\rm R}^{(k)}\;.
\end{eqnarray}
Here $[-\Delta + \nu_n^2 + {\mathcal{U}}_i(0,T)]^{-1}$ is a formal
representation of the Green's function associated to that operator.
The superscript $(k)$ denotes the order of the corresponding graph 
(see fig.~1).

In analogy also the improved one-loop effective potential
$V_{i,\rm R}(\phi,T)$ exists. It has the following relation to
the usual one-loop effective action $V_{i}(\phi,T)$:
\begin{eqnarray}
  V_{i,\rm R}(\phi,T)&=& c_i T \; {\rm tr} \; \sum_{n=-\infty}^{+\infty}
  \int \! \frac{{\rm d}^3p}{(2\pi)^3} \,
  \ln \left[ \mbox{\boldmath$p$\unboldmath}^2 + \nu_n^2 
    + {\cal U}_i(\phi,T)\right]\\
  &=& c_i T \; {\rm tr} \; \sum_{n=-\infty}^{+\infty}
  \int \! \frac{{\rm d}^3p}{(2\pi)^3} \,
  \ln \left[ \mbox{\boldmath$p$\unboldmath}^2 + \nu_n^2 
    + {\cal U}_i(\phi)\right]\\
  && + c_i T \; {\rm tr} \; \sum_{n=-\infty}^{+\infty}
  \int \! \frac{{\rm d}^3p}{(2\pi)^3} \,
  \ln \left[ \frac{\mbox{\boldmath$p$\unboldmath}^2 + \nu_n^2 
      + {\cal U}_i(\phi,T)}
    {\mbox{\boldmath$p$\unboldmath}^2 + \nu_n^2 
      + {\cal U}_i(\phi)}\right]\nonumber\\
  &=& V_i(\phi,T) + V_{i,\rm ring}(\phi,T)\;.
\end{eqnarray}
The divergencies in $V_i$ are
temperature independent. They are cancelled by the familiar
$T=0$--counter terms. However, also $V_{i,\rm ring}$ contains
divergencies which furthermore
are temperature dependent. Hence, the substitution of $U(\phi)$ by 
$V_{\rm ht}(\phi)$ accounts for finite and infinite contributions of
higher orders. The latter are removed
by the use of thermal counter terms \cite{Parwani:1992,Arnold:1993}. 
They are not fixed by renormalization conditions but are chosen
in such a way that they cancel the two divergent orders of 
$V_{i,\rm ring}(\phi,T)$ exactly. These two orders ($k=1$ and $k=2$)
can be read of from:
\begin{eqnarray}
  V_{i,\rm ring}(\phi,T)&=& \sum_{k=1}^{+\infty} V_{i,\rm ring}^{(k)} =
  c_i T \; {\rm tr} \; \sum_{n=-\infty}^{+\infty}
  \int \! \frac{{\rm d}^3p}{(2\pi)^3} \,
  \sum_{k=1}^{+\infty} \frac{(-1)^{k+1}}{k}\\
  && \times \left\{ 
    \left[ \frac{{\cal U}_i(\phi,T) 
        - {\cal U}_i(0,T)}{\mbox{\boldmath$p$\unboldmath}^2 
        + \nu_n^2 + {\cal U}_i(0,T)}
    \right]^k - \left[ \frac{{\cal U}_i(\phi) 
        - {\cal U}_i(0,T)}{\mbox{\boldmath$p$\unboldmath}^2 
        + \nu_n^2 + {\cal U}_i(0,T)}\right]^k\right\}\nonumber
\end{eqnarray}
However, the high temperature contribution of $V_{i,\rm ring}^{(2)}$, 
i.~e. the contribution of the static Matsubara mode, is finite
and is not subtracted.

For the numerical application the resummed one-loop effective action
is separated into a finite
part that is evaluated numerically and an infinite contribution that
has to be renormalized. The latter is given by the first two orders of
the series (\ref{sirphitexpansion}). In performing the renormalization
by adding the counter terms one finds contributions already taken into account
in the high temperature potential $V_{\rm ht}(\phi,T)$;
we denote these terms as $V_{i,\rm ht}$. These terms must be 
subtracted in order to avoid double counting; we call the resulting
difference $\Delta S_{i,\rm R,ren}^{(1+2)}$:
\begin{equation}
  \Delta S_{i,\rm R,ren}^{(1+2)}= S_{i,\rm R}^{(1)} + S_{i,\rm R}^{(2)} - 
  S_{i,\rm R,ht}^{(2)} - \beta\int \! {\rm d}^3x \,
  \left(
    V_{i,\rm ring}^{(1)} + V_{i,\rm ring}^{(2)}- V_{i,\rm ring,ht}^{(2)}
    +V_{i,\rm ht}
  \right)
  +S_{i,\rm ct}\;.
\end{equation}
Here we make use of that the high temperature approximation of the
second order contribution, $S_{i,\rm R,ht}^{(2)}$, is finite. Therefore,
it is treated together with the finite part, 
$S_{i,\rm R}^{\overline{(3)}}$, and is evaluated numerically:
\begin{equation}
  S_{i,\rm R,num}=S_{i,\rm R,ht}^{(2)} + 
  S_{i,\rm R,ht}^{\overline{(3)}} + S_{i,\rm
    R,ns}^{\overline{(3)}}
  =S_{i,\rm R,ht}^{\overline{(2)}} + S_{i,\rm
    R,ns}^{\overline{(3)}}\;.
\end{equation}
Here a superscript $\overline{(k)}$ denotes the sum of contributions
beginning with the $k$-th order. The 
index `ns' denotes the contributions of the
non-static Matsubara modes. It is known to be approximated
sufficiently by a gradient expansion \cite{Baacke:1996}. It is
found to be negligible small compared to the contribution due to the
static mode. Therefore it is neglected and only
$S_{i,\rm R,ht}^{\overline{(2)}}$ is evaluated numerically. This is postponed
to the next section.

Using
\begin{eqnarray}
  T \sum_{n=-\infty}^{+\infty} \int \! \frac{{\rm d}^3p}{(2\pi)^3} \,
  \frac{1}{\mbox{\boldmath$p$\unboldmath}^2 + \nu_n^2 + m^2}&=&
  \int \! \frac{{\rm d}^4p}{(2\pi)^4} \,
  \frac{1}{p^2+m^2}\\
  &&+\frac{T^2}{12}-\frac{mT}{4\pi}-\frac{m^2}{16\pi^2}\ln \frac{m^2}{a_BT^2}
   + \delta_1(T,m^2)\;,\nonumber\\
  T \sum_{n=-\infty}^{+\infty} \int \! \frac{{\rm d}^3p}{(2\pi)^3} \,
  \frac{1}{(\mbox{\boldmath$p$\unboldmath}^2 + \nu_n^2 + m^2)^2}&=&
  \int \! \frac{{\rm d}^4p}{(2\pi)^4} \,
  \frac{1}{(p^2+m^2)^2}\\
  +\frac{T}{8\pi m} + \frac{1}{16\pi^2}\ln \frac{m^2}{a_BT^2} + \delta_2(T,m^2)
  \nonumber
\end{eqnarray}
and neglecting the terms with $\delta_1$ and $\delta_2$, which are small
at the temperature we are working at,
the analytic part $\Delta S_{\rm R,ren}^{(1+2)}$ is found to be:
\begin{eqnarray}
  \label{shrana}
  \Delta S_{\rm R,ren}^{(1+2)}&=&\beta \int \! {\rm d}^3x \,
  \left\{
  \left[ \frac{3\lambda}{32\pi^2}
    \left( \frac{m_H^2}{2} \left( \ln \frac{m_H^2}{a_B T^2} + 2 \right) 
      - \check m_H^2\right)
    -\frac{3\lambda \check m_H T}{8\pi}
  \right]\phi^2 \right.\nonumber\\
  &&+\frac{g^3 T}{32\pi} \phi^3 +
  \left( \frac{g^2}{4} + \lambda\right)
  \left[ -\frac{3 \check m_H T}{8\pi} - \frac{3\check m_H^2}{32\pi^2}
    +\frac{3m_W^2}{32\pi^2}
    +\frac{3m_H^2}{64\pi^2} \ln \frac{m_W^2}{a_BT^2}\right] \phi^2\nonumber\\
  &&\left.
    +\frac{g^3T}{16\pi}\phi^3
    -\frac{3g^2}{32\pi^2} \left( \mbox{\boldmath$\nabla$\unboldmath} 
      \phi\right)^2
    \ln \frac{m_W^2}{a_BT^2}
  \right \}\;.
\end{eqnarray}
%
%
\section{Description of the numerical method}
\label{sec:numericmethods}
\subsection{The functional derivative of the effective action}
In the previous section the calculation of the
finite contribution to the resummed effective action
due to the different fluctuating fields was postponed. Combining them to
$S_{\rm R,ht}^{\overline{(2)}}$, its functional derivative with
respect to the background field $\phi$ can be
represented as
\begin{eqnarray}
  \label{funcder_genform}
  \frac{\delta S_{\rm R,ht}}{\delta \phi(\mbox{\boldmath$x$\unboldmath})}
  &=&{\rm tr} 
  \left[ 
    c \; \frac{\delta {\cal U}(\phi,T)}{\delta 
      \phi(\mbox{\boldmath$x$\unboldmath})}
    G(\mbox{\boldmath$x$\unboldmath},\mbox{\boldmath$x$\unboldmath})
  \right]\nonumber\\
  &=&{\rm tr} 
  \left[
    c \; \frac{\delta {\cal M}^2(\phi,T)}{\delta\phi(
      \mbox{\boldmath$x$\unboldmath})}G(\mbox{\boldmath$x$\unboldmath},
      \mbox{\boldmath$x$\unboldmath})
  \right]
  -g \; {\rm tr}
  \left[
    c \; \mbox{\boldmath$\zeta$\unboldmath} 
    \mbox{\boldmath$\nabla$\unboldmath} 
    G(\mbox{\boldmath$x$\unboldmath},\mbox{\boldmath$x$\unboldmath})
  \right]\;.
\end{eqnarray}
where ${\cal U}$, ${\cal M}^2(\phi,T)$ and 
$\mbox{\boldmath$\zeta$\unboldmath}$ are the $(6
\times 6)$ generalisations of the matrices defined in (\ref{ua0eta})ff. at
finite temperature. They are given explicitly in
appendix~\ref{app:partialwave}. $c$
contains the factors of degeneracy for the different fluctuations. Its
non-zero components are given by
\begin{equation}
  c_{11}=c_{22}=c_{33}=c_{44}=-c_{66}=\frac{3}{2}\;,\mbox{and} 
  \quad \quad c_{55}=\frac{1}{2} \;.
\end{equation}
The Green's function $G(\mbox{\boldmath$x$\unboldmath},
\mbox{\boldmath$x^\prime$\unboldmath})$ is the solution of 
\begin{equation}
  \biggl[ -\Delta + {\cal U}(\phi,T) \biggr] 
  G(\mbox{\boldmath$x$\unboldmath},\mbox{\boldmath$x^\prime$\unboldmath})= 
  \mbox{\boldmath$1$\unboldmath} \cdot 
  \delta^{(3)}(\mbox{\boldmath$x$\unboldmath}
  -\mbox{\boldmath$x^\prime$\unboldmath})\;.
\end{equation}
A method for the numerical computation of such a one-loop Green's function
was derived in \cite{Baacke:1990a,Baacke:1990b,Baacke:1992}.
In order to determine $G(\mbox{\boldmath$x$\unboldmath},
\mbox{\boldmath$x$\unboldmath})$ it is decomposed into its partial
wave contributions. The details of this calculation are given in
appendix~\ref{app:partialwave}. 
It turns out that by the decomposition one channel
of the coupled system can be combined with the Faddeev-Popov channel
so that the system reduces to a $(5 \times 5)$--system. 
Extracting the order $\overline{(2)}$ from (\ref{funcder_genform}) it
takes the form:
\begin{eqnarray}
  \label{funcder_result}
  \frac{\delta S_{\rm R,ht}^{\overline{(2)}}}{\delta\phi(
    \mbox{\boldmath$x$\unboldmath})}&=&
  \sum_{l=0}^{+\infty} \frac{2l+1}{4\pi} 
  \Biggl \{
    \frac{3g^2}{4}\phi\left( g_{11,l}^{\overline{(1)}} + 
      g_{22,l}^{\overline{(1)}}\right) 
    + \frac{3}{2}\left(\frac{g^2}{2}\phi + 2\lambda_T\phi-3ET\right)
    g_{33,l}^{\overline{(1)}}
  \nonumber\\
  && 
  + 3 (\lambda_T \phi -ET ) g_{44,l}^{\overline{(1)}}
  \Biggr \} -\frac{3g^2}{16\pi}\phi g_{55,0}^{\overline{(1)}}\\
  &&- \frac{3g}{2} \sum_l \frac{2l+1}{2\pi} \left\{ -c_1 \left(
      g_{13,l}^{\prime \overline{(1)}}
      +\frac{2}{r}g_{13,l}^{\overline{(1)}}\right) + c_0 
    \left( g_{23,l}^{\prime \overline{(1)}}+
      \frac{2}{r}g_{23,l}^{\overline{(1)}}\right)
  \right\}\;.\nonumber
\end{eqnarray}
Here the radial Green's functions $g_{km,l}(r,r)$ are defined by:
\begin{eqnarray}
  \label{gkmdefequ}
  && {\bf M}_{nk} \, g_{km,l}(r,r^\prime) =
  \frac{\delta_{nm}}{r^2}\delta(r-r^\prime)\;,\nonumber\\
  \mbox{with } &~& {\bf M}:={\bf M}^0 + V\;,\nonumber\\
  && {\bf M}^0= -\frac{{\rm d}^2}{{\rm d} r^2} - \frac{2}{r} \frac{{\rm
      d}}{{\rm d}r} + \frac{l_n(l_n+1)}{r^2} + m_n^2\;,\\
  &&m_n=(0 \;,\; 0 \;,\; \check m_H \;,\; \check m_H
  \;, 0)_n \;,\nonumber\\
  &&l_n=(l+1 \;,\; l-1 \;,\; l \;,\; l \;,\; l)_n\;.\nonumber
\end{eqnarray}
The non-zero components of the potential $V(\phi)$ are
\begin{eqnarray}
  V_{11}&=&V_{22}=V_{55}=\frac{g^2}{4}\phi^2(r) \;, \quad V_{33}=\left(
    \frac{g^2}{4} +\lambda_T \right) \phi^2(r) - 3ET\phi(r) \;,\nonumber\\
  V_{44}&=&3\lambda_T \phi^2(r) - 6ET\phi(r)\;, \quad
  V_{13}=V_{31}=-\sqrt{\frac{l+1}{2l+1}}g\phi^\prime(r) \;,\\
  \mbox{and} \quad
  V_{23}&=&V_{32}=\sqrt{\frac{l}{2l+1}}g\phi^\prime(r)\;.
  \nonumber
\end{eqnarray}
This coincides with \cite{Baacke:1995} where it
was derived by reduction of the sphaleron system.
The radial Green's function is given by \cite{Baacke:1990b}
\begin{equation}
\label{g_represent}
g_{km,l}(r,r^\prime)= \kappa \left[ f_{k,l}^{\alpha +}(r)
f_{m,l}^{\alpha -}(r^\prime)
\Theta(r-r^\prime) + f_{k,l}^{\alpha -}(r)
f_{m,l}^{\alpha +}(r^\prime)
\Theta(r^\prime-r) \right] \;,
\end{equation}
where $f_{k,l}^{\alpha +}$ and $f_{k,l}^{\alpha -}$ are the linear independent
solutions of the homogeneous equation
\begin{equation}
  \left[\left(\frac{\partial^2}{\partial r^2}+\frac{2}{r}
      \frac{\partial}{\partial r} - \frac{l_n(l_n+1)}{r^2} - m_n^2
    \right) \delta_{nk}
    -V_{nk}(\phi,T)\right] f_{k,l}^{\alpha \pm}(r)=0\;.
\end{equation}
The mode functions $f_{k,l}^{\alpha +}$ ($f_{k,l}^{\alpha -}$)
are regular (singular) at infinity. In order to calculate the
mode functions numerically their asymptotic behaviour at infinity is
separated from the rest by the following ansatz \cite{Baacke:1992}:
\begin{equation}
f_{k,l}^{\alpha\pm}(r) = b_{l_k}^\pm(z)\left( 
\delta_k^\alpha+h_{k,l}^{\alpha\pm}(r)\right) \qquad \mbox{with} \qquad 
z=\kappa r\;.
\end{equation}
The functions $b_{l_n}^\pm(z)$ satisfy the equation 
\begin{equation}
\left(\frac{\partial^2}{\partial z^2}+\frac{2}{z}
\frac{\partial}{\partial z} - \frac{l_n(l_n+1)}{z^2} - 1\right)
b_{l_n}^\pm(z)=0\;,\\
\end{equation}
One obtains
\begin{equation}
\label{eq:besselfunc}
\begin{array}{ll}
\displaystyle
b_{l_1}^+(z)= z^{-(l+2)}(2l+1)!! &
\displaystyle b_{l_1}^-(z)=\frac{z^{l+1}}{(2l+3)!!}\,,\\
\displaystyle b_{l_2}^+(z)= z^{-l}\frac{(2l+1)!!}{(2l-1)(2l+1)} &
\displaystyle b_{l_2}^-(z)=\frac{z^{l-1}(2l+1)}{(2l+1)!!}\,,\\
\displaystyle b_{l_3}^+(z)=b_{l_4}^+(z)=k_l(z) &
\displaystyle b_{l_3}^-(z)=b_{l_4}^-(z)=i_l(z)\;,\\
\displaystyle b_{l_5}^+(z)= \frac{1}{z}&
\displaystyle b_{l_5}^-(z)= 1\;.\\
\end{array}
\end{equation}
In the massive channels we obtain the modified Bessel functions, while
in the massless ones the solutions reduce to powers of $z$.
The fifth channel only contributes in the s--wave. Therefore, its Bessel
function is given only for $l=0$.
The normalization constants and $\kappa=\check m_H$ 
are chosen in such a way that the
Wronskians of the different channels become equal:
\begin{equation}
W\Bigl[ b_{l_n}^+(z),b_{l_n}^-(z)\Bigr]=\frac{1}{z^2}\,.
\end{equation}
The differential equation of the rest mode functions $h_n^{\alpha\pm}$ 
reduces to:
\begin{equation}
  \label{dglhnalpha}
  \left[\frac{\partial^2}{\partial r^2}+2 \left(\frac{1}{r}
      + \kappa \frac{b_{l_n}^{\prime \pm}(z)}{b_{l_n}^\pm(z)}\right)
    \frac{\partial}{\partial r} \right]
  h_{n,l}^{\alpha\pm}(r)=V_{nk}
  \left( \delta_k^\alpha+h_k^{\alpha\pm}(r) \right)
  \frac{b_{l_k}^\pm(z)}{b_{l_n}^\pm(z)}\;.
\end{equation}
This equation can be used for generating the functions
$h_{n,l}^{\alpha \pm}$ order by order in the potential
$V$. Introducing the contribution of order $k$ in the potential as
$h_{n,l}^{\alpha \pm (k)}$ and defining
\[h_{n,l}^{\alpha \pm
  \overline{(k)}}:= \sum_{j=k}^{+\infty}h_{n,l}^{\alpha \pm (j)}
\]
as in \cite{Baacke:1994}, the order $\overline{(1)}$ of the Green's
function is given by:
\begin{eqnarray}
  \label{calcghl}
  g_{km,l}^{\overline{(1)}}(r,r)&=&
  \frac{\kappa}{2} 
  \left[
    b_{l_m}^-(z) b_{l_k}^+(z) 
    \left( 
      h_{m,l}^{k - \overline{(1)}}(r) + h_{k,l}^{m + \overline{(1)}}(r)
      + h_{k,l}^{\alpha + \overline{(1)}}(r) h_{m,l}^{\alpha -
        \overline{(1)}}(r)
    \right)
  \right.\nonumber\\
  &&
  \left. + b_{l_k}^-(z) b_{l_m}^+(z)
    \left(
      h_{k,l}^{m - \overline{(1)}}(r) + h_{m,l}^{k + \overline{(1)}}(r)
      + h_{m,l}^{\alpha + \overline{(1)}}(r) h_{k,l}^{\alpha -
        \overline{(1)}}(r)
    \right)
  \right]
\end{eqnarray}
%
%
\subsection{Evaluation of the effective action}
In order to calculate the nucleation rate $\gamma$
the effective action in the one-loop
approximation is needed. A method to calculate its functional derivative
with respect to the background field 
has been given in the previous subsection. For the purpose of renormalization
the divergent parts have been separated from the rest and have been handled 
analytically. The same separation can be used here in order to 
evaluate the effective action itself. The analytic part is given
by (\ref{shrana})ff. The numerical contributions 
$S_{i,\rm ht}^{\overline{(2)}}$ are determined in the following.

For the numerical computation of the effective action the fluctuation
determinant ${\cal J}$ is decomposed into partial waves 
\begin{eqnarray}
  \ln {\cal J} &=& \sum_{l} (2l+1) \ln {\cal J}_l\;,\\
  \mbox{with} \quad {\cal J}_l &=& \frac{\det {\bf M}}{\det {\bf M}^0}
  \;.\nonumber
\end{eqnarray}
Here ${\bf M}$ and ${\bf M}^0$ are the fluctuation operators defined in
(\ref{gkmdefequ}). 
We need to know two $(n \times n)$ matrices $\widetilde {\bf f}_l(\nu,r)$ and 
$\widetilde {\bf f}_l^0(\nu,r)$
which contain the $n$ linear independent solutions
$\widetilde f_{k,l}^\alpha(\nu,r)$ and $\widetilde f_{k,l}^{\alpha 0}(\nu,r)$
of the system of coupled differential equations
\begin{equation}
\label{mnkfka0}
\left( {\bf M} + \nu^2 \right)_{nk} \widetilde f_{k,l}^\alpha(\nu,r) =0 \qquad
\mbox{and} \qquad
\left( {\bf M}^0 + \nu^2 \right)_{nk} \widetilde f_{k,l}^{\alpha
  0}(\nu,r) 
=0\;.
\end{equation}
The boundary conditions of these functions are chosen to be regular
at $r=0$. The lower index denotes the $n$ components while the
different solutions are labelled by the Greek upper index. 
These solutions are normalized such that
\begin{equation}
\lim_{r\to 0} \widetilde f_l(\nu,r) 
\left( \widetilde f_l^0(\nu,r)\right)^{-1}=\mbox{\boldmath$1$\unboldmath}
\;.
\end{equation}
On these conditions the statement of the
theorem is \cite{Baacke:1994}:
\begin{equation}
  {\cal J}_{i,l}(\nu):=\frac{\det \left( {\bf M} + \nu^2 \right)}{\det
    \left({\bf M}^0 + \nu^2\right)} 
  = \lim_{r \to \infty} \frac{\det \widetilde f(\nu,r)}
  {\det \widetilde f^0(\nu,r)}\;.
\end{equation}
The theorem has been applied for computing the one-loop effective
action of a single scalar field on a bubble background
\cite{Baacke:1993}, and of a fermion system on a similar background
\cite{Baacke:1994} previously. Furthermore it has been used to
calculate the bosonic fluctuation determinant of the critical bubble
\cite{Baacke:1995}.
It was found to yield very precise
results, in addition to providing a very fast computational method.

If the theorem is applied at $\nu=0$ the fluctuation determinant
${\cal J}\equiv{\cal J}(0)$ is obtained. The consideration of finite
values of $\nu$ is necessary in the discussion of zero modes.
For simplicity the following formulas are given only for $\nu=0$.

In analogy to the mode functions $f_l^\pm(r)$ 
also the functions $\widetilde f_l(r)$ can be decomposed as:
\begin{equation}
  \widetilde f_{k,l}^\alpha(r)= b_{l_k}^-(\kappa r) 
  \Bigl( \delta_k^\alpha+ \widetilde
  h_{k,l}^\alpha(r)\Bigr)\;,
\end{equation}
with $b_{l_k}^-(\kappa r)$ given by (\ref{eq:besselfunc}).
The consequence is that the separated functions $\widetilde h_k^\alpha(r)$
satisfy the same type of differential equation as the rest mode functions
$h_{k,l}^{\alpha -}(r)$. Only the boundary conditions differ,
namely $\widetilde h_{k,l}^\alpha(0)=0$
and $\widetilde h_{k,l}^\alpha(\infty)=\mbox{const}$. The two
functions are related by:
\begin{equation}
h_l^-(r) = \left[ \mbox{\boldmath$1$\unboldmath}+\widetilde h_l(r)\right] 
\left[ \mbox{\boldmath$1$\unboldmath}
+\widetilde h_l(\infty) \right]^{-1} - \mbox{\boldmath$1$\unboldmath}\;.
\end{equation}
After determining $\widetilde f_l(r)$ one obtains the desired
order of the fluctuation determinant.
\begin{equation}
\ln {\cal J}_l^{\overline{(2)}}=
\lim_{r\to\infty} \left\{\ln \det [1+\widetilde h_l(r)] - {\rm tr} \,
\widetilde h_l^{(1)}\right\}\;.
\label{j2bar}
\end{equation}
As the determinant ${\cal J}$ is a product of three individual
determinants due to the coupled and uncoupled channels one has to
take into account the different factors of degeneracy for each of them
(see (\ref{funcder_genform})).
%
%
\subsection{Self-consistent determination of the bubble profile}
\subsubsection{Computing the start profile for the iteration}
\label{sec:startprofile}
In order to solve 
the one-loop field equation iteratively we have to choose a suitable
separation of the equation into a left-hand and right-hand
side. Then setting the right-hand side to zero results in a first
approximate solution of the equation. This is taken as the 
start profile for the iteration.

In particular we choose the following form of the field equation
\begin{equation}
  \label{fieldequation3}
  -\Delta \phi + \frac{\partial}{\partial \phi} V_{\rm R}(\phi,T)
  = {\cal I}(\phi)\;,
\end{equation}
with
\begin{eqnarray}
  V_{\rm R}(\phi,T)&:=&V_{\rm ht}(\phi,T) 
  - \frac{T}{12\pi} \left( 3\lambda_T\phi^2 + \check m_H^2 \right)^\frac{3}{2}
  + \frac{3\lambda_T \check m_H T \phi^2}{8\pi} + \frac{\check m_H^3 T}{12\pi}
  \nonumber\\
  &&-\frac{T}{4 \pi} \left[
    \left( \frac{g^2}{4}+\lambda_T \right) \phi^2 - 3ET\phi 
    + \check m_H^2\right]^\frac{3}{2}
  +\frac{3\check m_H T}{8\pi}
  \left[
    \left(
      \frac{g^2}{4}+\lambda_T
    \right)\phi^2 - 3ET\phi
  \right]\nonumber\\
  &&+ \frac{\check m_H^3 T}{4\pi} + \frac{g^3T}{32\pi}\phi^3
  +\left[
    -\frac{3\lambda\check m_H T}{8\pi} + \frac{3\lambda}{32\pi^2}
    \left(
      \frac{m_H^2}{2}
      \left(
        \ln \frac{m_H^2}{a_BT^2}+2
      \right)-\check m_H^2
    \right)
  \right]\phi^2\nonumber\\
  &&+
  \left(
    \frac{g^2}{4}+\lambda
  \right)
  \left[
    -\frac{3\check m_H T}{8\pi} - \frac{3(\check m_H^2 - m_W^2)}{32\pi^2}
    +\frac{3m_H^2}{64\pi^2} \ln \frac{m_W^2}{a_BT^2}
  \right]\phi^2
\end{eqnarray}
I.~e., in addition to $V_{\rm ht}$ the contributions of the
isoscalar and isovector part of the Higgs field 
to the resummed effective potential are
shifted to the left-hand side. For a suitable chosen temperature
the potential $V_{\rm R}$ has a secondary minimum at a nonzero value of $\phi$.
We denote this minimum as $\widetilde v(T)$. It is used to
scale the dimensional quantities as follows:
\begin{equation}
  \phi=\widetilde v(T) \bar \phi\;, \quad
  x_\mu=\frac{1}{g\widetilde v(T)} \bar x_\mu\;, \quad \mbox{and} \quad
  T=g\widetilde v(T) \bar T\;.
\end{equation}
All masses are scaled with $g\widetilde v(T)$.
The inhomogeneity ${\cal I}(\phi)$ consists of (\ref{shrana}) 
and (\ref{funcder_result}) reduced by those
terms that were shifted to the left-hand side.
\begin{eqnarray}
  \label{inhomogeneity}
  {\cal I}(\phi)
  &=&-\frac{3T}{2}\left\{ \frac{g^2}{2}\phi 
    \sum_{l=0}^{+\infty} \frac{2l+1}{4\pi}
    \left( g_{11,l}^{\overline{(1)}} + g_{22,l}^{\overline{(1)}}
    \right)+\left(\frac{g^2}{2}\phi+2\lambda_T\phi-3ET
    \right) \sum_{l=0}^{+\infty}
    \frac{2l+1}{4\pi} g_{33,l}^{\overline{(1)}}\right\}\nonumber\\
  &&+\frac{3gT}{2}\sum_{l=0}^{+\infty} \frac{2l+1}{2\pi}
  \left[ -c_1\left( g_{13,l}^{\prime\overline{(1)}}+\frac{2}{r}
      g_{13,l}^{\overline{(1)}}\right)
    +c_0\left( g_{23,l}^{\prime\overline{(1)}}
      +\frac{2}{r}g_{23,l}^{\overline{(1)}}\right)\right]\nonumber\\
  &&-3T\left( \lambda_T\phi -ET\right) \sum_{l=0}^{+\infty} \frac{2l+1}{4\pi}
  g_{h,l}^{\overline{(1)}} 
  +\frac{3g^2T}{16\pi}\phi g_{a_0\eta,0}^{\overline{(1)}}\\
  &&-\frac{3\lambda_T T \phi}{4\pi}
  \left[
    \left(
      3\lambda_T\phi^2 + \check m_H^2
    \right)^\frac{1}{2}-\check m_H
  \right]-\frac{3g^3T}{16\pi}\phi^2 - \frac{3g^2}{32\pi^2} \Delta \phi \ln
  \frac{m_W^2}{a_BT^2}\nonumber\\
  &&-\frac{3T}{8\pi}
  \left[
    2
    \left(
      \frac{g^2}{4}+\lambda_T
    \right)\phi-3ET
  \right]
  \left[
    \left(
      \left(
        \frac{g^2}{4}+\lambda_T
      \right)\phi^2 - 3ET\phi+\check m_H^2
    \right)^\frac{1}{2}-\check m_H
  \right]\nonumber\;.
\end{eqnarray}
For the purpose of comparison the solution $\phi_{\rm cl}$
of the equation
\begin{equation}
  \label{classprofequ}
  -\Delta \phi_{\rm cl} 
    + \left. \frac{\partial}{\partial \phi}
        V_{\rm ht}(\phi,T)
    \right\vert_{\phi=\phi_{\rm cl}}=0
\end{equation}
is also calculated. It is denoted as the classical profile.

In order to avoid large changes between two subsequent profiles 
especially at the beginning of the
iteration we use a relaxation, i.~e. the field equation 
(\ref{fieldequation3}) is written as 
\begin{equation}
  -\Delta \phi^{(n+1)} 
  + \left.\frac{\partial}{\partial \phi}
    V_{\rm R}(\phi,T)\right\vert_{\phi=\phi^{(n+1)}}=
  \left( 1-\varepsilon \right) {\cal I}(\phi^{(n-1)}) 
  + \varepsilon {\cal I}(\phi^{(n)})\;.
\end{equation}
By a suitable chosen $\epsilon$ the convergence of
the iterations of the profile can be improved.
%
%
\subsubsection{Treatment of zero modes and unstable mode}
For the isoscalar fluctuations there are two contributions to the
effective action and its functional derivative that have to be treated
individually: namely the instable mode in the partial wave $l=0$ and
the translation zero mode in the partial wave $l=1$.

A bound state occurs in the s-wave, i.~e.\ a state with a negative
eigenvalue. As this is the only negative eigenvalue the total
determinant is also negative and the one-loop contribution to the
effective action is complex. This corresponds to the fact that the
state under consideration, the critical bubble, is unstable.

In the p-wave there exists a mode with the eigenvalue zero. It
corresponds to the fact that the critical bubble breaks
translational invariance. Because of this mode the
determinant itself is zero and the effective action diverges.

For the calculation of the nucleation rate instead of the fluctuation
determinant with these two properties we need one where the negative
eigenvalue has been replaced by its modulus and where the three
eigenvalues zero have been removed. These changes to the determinant
are indicated by two primes:
\begin{equation}
{\cal J}_h= \prod_{n=-\infty}^{+\infty}
\frac{\det { }^{\prime\prime} \left[-\Delta +\nu_n^2 + m_h^2(\phi,T)
\right]}{\det \left[ -\Delta
+\nu_n^2 + m_h^2(0,T) \right]}\;.
\end{equation}
As three eigenvalues zero have been removed from the
determinant of the isoscalar fluctuations the rate gets its dimension
$({\rm energy})^3$, i.~e.\ $({\rm volume})^{-1}$.
The further dimension $({\rm time})^{-1}$ is due to the eigenvalue
$\omega_-$ of the unstable mode.

The numerical method presented in the previous sections does not 
explicitly make use of the eigenvalues. 
Therefore, we have to show how these
two contributions are removed from the effective action
-- this has also been discussed in \cite{Baacke:1995} -- and its
functional derivative.

In case of the unstable mode this is performed easily. One
computes the particular partial wave contributions according to
(\ref{j2bar}). In the partial wave $l=0$ only the sign of the
determinant has to be inverted before taking the logarithm. For the
functional derivative the sign is irrelevant because due to
\begin{equation}
\frac{\delta}{\delta \phi(\mbox{\boldmath$x$\unboldmath})} \ln {\cal J}=
\frac{\delta}{\delta \phi(\mbox{\boldmath$x$\unboldmath})} \ln 
\prod_{\alpha} \omega_\alpha^2 =
\sum_{\alpha} \frac{1}{\omega_\alpha^2} \frac{\delta
  \omega_\alpha^2}{\delta \phi(\mbox{\boldmath$x$\unboldmath})}
\end{equation}
the sign of each eigenvalue enters the equation twice.

The removal of the translational zero mode is more complicated.
An alternative representation for the Green's function $g_{44,l}(r,r^\prime)$
compared to (\ref{g_represent}) can be deduced from the eigenvalues and
eigenfunctions of
\begin{equation}
\left[-\frac{\partial^2}{\partial r^2} 
- \frac{2}{r} \frac{\partial}{\partial r}
+\frac{l(l+1)}{r^2} + \check m_H^2 + 3 \lambda_T\phi^2-6ET\phi \right]
\psi_\alpha(r)=\omega_\alpha^2 \psi_\alpha(r)\;.
\end{equation}
It is given by:
\begin{equation}
g_{44,l}(r,r^\prime)= \sum_\alpha 
\frac{\psi_\alpha(r)\psi_\alpha^\ast(r^\prime)}{\omega_\alpha^2}\;.
\end{equation}
Obviously this Green's function is divergent when the spectrum of the
operator contains the eigenvalue zero. Defining a new Green's function
as
\begin{equation}
g_{44,l}(r,r^\prime,\nu)= \sum_\alpha 
\frac{\psi_\alpha(r)\psi_\alpha^\ast(r^\prime)}
{\nu^2 + \omega_\alpha^2}
\end{equation}
the divergence manifests itself in the limit $\nu\to 0$ 
in the form of a $1/\nu^2$--pole.
As the zero mode is the only mode which contributes to this pole its
contribution can be removed by defining the Green's function at
$\nu=0$ without the zero mode in the following way:
\begin{equation}
g_{44,l}(r,r^\prime,0):=\lim_{\nu\to 0} \left[ g_{44,l}(r,r^\prime,\nu)
-\frac{\psi_0(r)\psi_0^\ast(r^\prime)}{\nu^2}\right]\;.
\end{equation}
In the special case under consideration, where the Green's function
only has to be calculated at $r=r^\prime$, the practical way to
subtract the zero mode in the p--wave is to compute the function 
$g_{44,1}(r,r,\nu)$ for sufficient many different values of $\nu$.
Then in the function
\begin{equation}
\hat g_{44,1}(r,r,\nu):= \frac{A(r)}{\nu^2} + B(r) + C(r)\nu^2
\end{equation}
the parameters $A$, $B$ and $C$ are adjusted by a fit with respect
to $\nu$ for each value of $r$. The subtraction of the zero mode
is then done by setting $g_{44,1}(r,r,0):=B(r)$.

The computation of the function $g_{44,1}(r,r,\nu)$ is done in the same
way as described in section~\ref{sec:numericmethods}. The only
difference is, in the definition of the radial part of the Green's
function (\ref{gkmdefequ}) and in all consecutive equations a $\nu^2$ 
has to be added to the partial wave operator ${\bf M}^0$.
Hence, the `mass' $\kappa$ of the isoscalar channel
changes to $\kappa=\sqrt{\check m_H^2 + \nu^2}$.

The one-loop contribution of the isoscalar fluctuation to the
fluctuation determinant
is determined according to (\ref{j2bar}) in each partial wave.
In the p--wave the following expression has to be evaluated:
\begin{equation}
\lim_{r\to\infty} \ln \left[ 1+\widetilde h_{4,1}^4(r)\right]\;.
\end{equation}
The numerical investigation shows that $\widetilde h_{4,1}^4(\infty)=-1$
and the logarithm diverges due to the
zero mode in this partial wave.
In order to subtract the mode the computational method to determine the
fluctuation determinant is extended in the same way as before in
case of the Green's function. I.~e., the theorem mentioned
above is applied at finite $\nu$.
In order to subtract the zero mode the
determinant ${\cal J}_{h,1}(\nu)$ has to be replaced by 
\cite{Baacke:1995,Baacke:1993}
\begin{equation}
\lim_{\nu\to 0} \lim_{r\to\infty}
\frac{{\rm d}}{{\rm d}(\nu^2)} \widetilde h_{4,1}^4(\nu,r)\;.
\end{equation}
In practice $\widetilde h_{4,1}^4(\nu,\infty)$ is computed
for some sufficient small values of $\nu$
and is used to fit the constants $A$, $B$ and $C$ in
\begin{equation}
1+\widetilde h_{4,1}^4(\nu,\infty)=A\nu^2 + B\nu^4 + C\nu^6\;.
\end{equation}
With these techniques the unstable mode and the translation mode are
removed from the effective action as well as from its functional
derivative. Concerning the translation mode a further remark is necessary.
An exact zero mode only exists when the profile $\phi$ is
determined from the classical field equation (\ref{classprofequ}).
The one-loop equation differs from the
classical one due to the non-vanishing right-hand side of the equation.
As a consequence one finds a bound state in the spectrum in the
p--wave below the continuum threshold but its eigenvalue only is near
to zero but it is not exact zero. Nevertheless this mode is subtracted.

In the coupled system of gauge fields and would-be-Goldstone bosons a zero
mode exists in the s-wave. For the treatment of this mode we
adpopt the prescription given in \cite{Baacke:1995}. 
%
%
\subsubsection{Some details of the numerical evaluation}
\label{numerik_sonstiges}
In order to determine the profile for one step of the iteration a boundary
problem has to be solved. At $r=0$ the slope of the profile has to
vanish and at $r=\infty$ the function has to tend to zero
exponentially. The profile is obtained using a shooting method: The
value of the profile at $r=0$ is varied until the profile gets the
correct behavior at infinity. We typically use 1000 steps on a
scale from $r=0$ to $r_{\rm max}\approx 1/\check m_H$ where the latter
typically is below 200 in the units $g \widetilde v$ we are working in.
The numerical technique used to integrate the differential
equation is the Nystr\"om method.

For small values of $r$ the profile can be represented as
$\phi(r)\approx a_0 + a_2 r^2 + a_4 r^4$. 
Hence, for a given profile the parameters can be obtained using a
simple fit. Then for the rest mode functions $\widetilde
h^{\overline{(1)}}$ which are calculated from the differential equations
(\ref{dglhnalpha})
the behavior at small $r$ can be evaluated analytically using a power
series expansion. Using this expansion the functions have the correct
starting properties for the following numerical integration. Hence, we
can avoid numerical problems due to the divergent behavior of the
Bessel functions in this region.

The radial Green's functions are composed from the
rest mode functions according to (\ref{calcghl})
and then the partial wave summation
(\ref{inhomogeneity}) is performed. This sum is executed explicitly up to
$l_{\rm max}=15$. At large $l$ the product $(2l+1) g_{i,l}(r,r)$
behaves as $a(r)/l^2 + b(r)/l^3 + c(r)/l^4$, no matter which channel
$i$ one considers. Using a fit the coefficients $a$, $b$ and $c$ can
be obtained for each value of $r$ and then the value of the sum from
$l_{\rm max}$ to infinity can be estimated. As a cross-check we start
this estimate at $l=10$ and it is tested whether it
approaches a constant when $l$ reaches $l_{\rm max}$.

In order to check the precision various cross-checks have been
implemented. The equality
\begin{equation}
  \label{wronski_coupled}
  \omega^{\beta\alpha}:=
  \check m_H r^2 W\left[ f_{n,l}^{\alpha+}(r), f_{n,l}^{\beta -}(r)
  \right] = \delta_\alpha^\beta
\end{equation}
is satisfied numerically with a relative deviation of the order
$O(10^{-6})$. Especially the Wronskian between $f^+(r)$ and
$\widetilde f(r)$ satisfies:
\begin{eqnarray}
  \label{wronski_ident1}
  \kappa r^2 W\left[ f^{\alpha+}(r),\widetilde f^\beta(r) \right]
  &=&\delta_\alpha^\beta + \widetilde h_\alpha^\beta(\infty)\\
  &=&\delta_\beta^\alpha + h_\beta^{\alpha+}(0)\;.
  \label{wronski_ident2}
\end{eqnarray}
The relative deviations from these identities are of the order
$O(10^{-4})$ for (\ref{wronski_ident1}) and $O(10^{-6})$ for
(\ref{wronski_ident2}). The former being not as precise as the latter
is due to the bad convergence of the functions
$\widetilde h_\alpha^\beta$ at large $r$. In the massless channels they only
achieve their asymptotic values as $1/r^l$. Hence, their value at
$r=\infty$ only can be estimated from a fit.
On the other hand the constancy of the product of $r^2$ with the
Wronskian is satisfied excellently. Therefore we do not make use of
this fitted quantity in the further process.
%
%
\section{Results and Discussion}
\label{sec:results}
The numerical results are given in table~1 and
in the figures~2 to 7. 
The Higgs boson masses used here are 30GeV and 40GeV.
The mass of the top-quark is taken to be $m_t=170$GeV 
and for the gauge boson mass we use a value of $m_W=80.2$GeV. The gauge
coupling $g$ is taken as $g=.651616$. The vacuum expectation value of the
Higgs field is then fixed to be $v_0=246$GeV.
The relevant temperature interval is determined from the approximation
of the resummed effective potential $V_{\rm R}$. It is found to be smaller
than the interval determined from the high temperature approximation
$V_{\rm ht}$ of the effective potential. The values are given in
table~1.
Within this interval we
choose equidistant steps for the temperature. We leave out the range
of temperatures close to the roll-over temperature $T_0$, 
because these correspond
to small thick wall bubbles which are of minor cosmological interest.

We give the results for the logarithm of the nucleation rate
$\gamma$, i.~e. the rate including the one-loop corrections and prefactors
as defined in (\ref{nuclrate}).
It is evaluated using two different bubble profiles.
First we use the self-consistent profile, i.~e. the solution of 
(\ref{fieldequation3}). The results have to be compared to that ones we obtain
from using the profile $\phi_{\rm cl}$, which is the solution
of (\ref{classprofequ}).
The results are displayed in the
figures~2 and 3 where
we give the logarithm of the nucleation rate versus the temperature.
For the results based on the self-consistent profile we also present
a function that fits the explicitly calculated points very well.
The function has the form
\begin{equation}
  \label{func:F_t}
  F(T)=-a \left( \frac{(T_c^{\rm R })^2}{(T_c^{\rm R})^2 - T^2}\right)^b\;.
\end{equation}
The results for $a$ and $b$ depend on the specific parameter set.
We find $a=3.48\cdot 10^{-2}$ and $b=1.350$ for $m_H=30$GeV and 
$a=2.38\cdot 10^{-2}$
and $b=1.362$ for $m_H=40$GeV.

At $m_H=30$GeV the self-consistently calculated transition rate is smaller
than the rate obtained from the classical profile, while at $m_H=40$GeV
it is just the other way around. This is due to the relation between
the temperature intervals. While in both cases the interval suitable for 
finding the self-consistent solution is located within the interval derived
from the high temperature potential it is shifted towards higher temperatures
with increasing Higgs boson mass (see table~1).
For comparison we map the temperature intervals $[T_0^{\rm R},T_c^{\rm R}]$
and $[T_0^{\rm ht},T_c^{\rm ht}]$ onto the interval $[0,1]$ by introducing
the dimensionless quantity $\zeta=(T-T_0)/(T_c-T_0)$. The results of
both parameter sets are
given in figure~4.
For both parameter sets
the self-consistently obtained rate is increased compared
to the ordinary one. Hence, the effect of using the self-consistent
profile is an enhancement of bubble nucleation.

Based on the results for the nucleation rate we can investigate some aspects
of the cosmological electroweak phase transition. As the universe expands
the temperature $T$ decreases. Once the critical temperature $T_c$ is reached
critical bubbles can start to nucleate and grow. The corresponding time
is obtained using the relation between time and temperature \cite{Kolb:1990}:
\begin{equation}
  t \approx \frac{0.03 m_{\rm PL}}{T^2} 
  \qquad \mbox{with } m_{\rm PL}=1.2\cdot10^{19}\mbox{GeV}\;.
\end{equation}
Hence, the nucleation rate $\gamma$ can be considered as a function of
time as well. For a given velocity  $v$ of the bubble wall
the number density of bubbles of size
$r$ at the time $t$ is given by
\begin{equation}
  n(t,r)=\gamma\left(t-\frac{r}{v}\right)\;.
\end{equation}
Then the fraction of space that is still in the symmetric phase is obtained
as \cite{Guth:1981uk}:
\begin{equation}
  \rho_s(t)=\exp
  \left[
    -\frac{4\pi}{3} \int \limits_{t_c}^{t}\! {\rm d} t_1 \,
    v^3(t-t_1)^3 \gamma(t_1)
  \right]\;.
\end{equation}
This quantity is used to define implicitly the
time $t_e$ when the nucleation is completed by $\rho_s(t_e)=1/e$.
As it marks the end of the phase transition it is a quantity that is easier
accessible in lattice simulations than other characteristics of the transition.
The fraction of space that is already in the asymmetric phase
is simply $\rho_a(t)=1-\rho_s(t)$.
The total number of droplets per unit volume, $N(t)$, also depends on
the fraction $\rho_s(t)$ because new bubbles can only nucleate in a region that
is still in the symmetric phase:
\begin{equation}
  N(t)=\int \limits_{t_c}^{t}\! {\rm d} t_1 \,
  \gamma(t_1) \rho_s(t_1)\;.
\end{equation}
This quantity is important for the discussion of structure formation.
The average bubble radius, $R(t)$, is obtained as:
\begin{equation}
  R(t) = \frac{1}{N(t)} \int \limits_{t_c}^{t}\! {\rm d} t_1 \,
  v (t-t_1)\gamma(t_1)\rho_a(t_1)\;.
\end{equation}
The quantities $\rho_a(t)$, $N(t)$ and $R(t)$ are evaluated
for the parameter set $m_H=30$GeV, $m_t=170$GeV for different values
of the bubble wall velocity $v$. The results are displayed in
figure~5 to 7. The completion time $t_e$
depends only weakly on the bubble wall velocity and is roughly 
$t_e-t_c \approx 0.15 \cdot 10^{-12}$sec (see figure~5). 
This can be compared to 
a previous calculation of this quantity
\cite{Bodeker:1994kj}, where a value of roughly $3 \cdot 10^{-12}$sec is
given. One sees the effect of enhancing
the rate. The time scale on which the nucleation proceeds is decreased
by an order of magnitude. Also for the number of bubbles per volume
after the completion time the 
results can be compared. For three values of the bubble wall velocity, namely
$c$, $c/10$ and $c/1000$ the corresponding values for $\ln(N/{\rm cm}^3)$ 
can be read of from figure~6: they are
$28.1$, $34.7$ and $47.8$. 
For the same velocities in \cite{Bodeker:1994kj}
the values $9.3$, $12.3$ and $18.0$ are given; hence, the 
number of bubbles per volume is increased substantially.
The average bubble radius $R(t)$ 
for the three velocities under consideration
comes out to be the same as in \cite{Bodeker:1994kj} and is plotted in
figure~7. As the initial radius of the critical bubble is
of the order of $1$fm it is neglected throughout this calculation.

The presented method to calculate self-consistently the solution of a one-loop
field equation has the following essential features:
\begin{itemize}
\item The complete one-loop determinant and its functional derivative is 
  evaluated. No expansion in any quantity is needed. In view of the
  fact that the derivative and heat kernel expansion generate
  unphysical infrared divergencies order by order this is an important
  conceptual advantage of the method presented here.
\item Coupled with an iterative procedure the numerical computation of
  the functional derivative of an effective action allows the
  determination of a self-consistent solution of a full one-loop field
  equation.
\item The divergent and convergent contributions to the effective
  action can be treated separately. While the convergent part is
  calculated numerically the divergent part is evaluated
  analytically. So the choice of the regularization and
  renormalization prescription is free.
\item The numerical computation of the functional derivative of an
  effective action only requires numerical integration of a few differential
  equations per partial wave. This can be done with high precision and
  is not computer time consuming.
\end{itemize}
In conclusion we find that taking into account the radiative corrections
already in the determination of the bounce solution has the following
effects:
\begin{itemize}
\item The width of the temperature interval within that a first order 
  phase transition is possible is decreased and with increasing the Higgs boson
  mass it is shifted towards higher temperatures.
\item The transition rate is increased substantially relative to the
  rate obtained from the standard calculation.
\item The time scale within that the transition is completed is reduced.
  The number of bubbles per volume is increased.
\end{itemize}
\section*{Acknowledgements}
It is a pleasure to thank J.~Baacke for useful discussions concerning
the numerical method and the treatment of the zero modes and for reading 
the manuscript.
%
%
\begin{appendix}
\section{Partial wave decomposition of the Green's function}
\label{app:partialwave}
In \cite{Baacke:1995} the partial wave decomposition of the Green's
function was done by reduction of the sphaleron system. An alternative
derivation will be given here.
The Green's function $G(\mbox{\boldmath$x$\unboldmath},
\mbox{\boldmath$x^\prime$\unboldmath})$ introduced in (\ref{funcder_genform})
satisfies the matrix equation
\begin{equation}
  \left( 
    -\frac{\partial^2}{\partial r^2} -
    \frac{2}{r}\frac{\partial}{\partial r} + \frac{1}{r^2}{\bf I\!L}^2 
    + {\cal M}^2(\phi,T) + g\mbox{\boldmath$\zeta$\unboldmath} 
    \mbox{\boldmath$\nabla$\unboldmath} \phi 
  \right) 
  G(\mbox{\boldmath$x$\unboldmath},\mbox{\boldmath$x^\prime$\unboldmath})= 
  \mbox{\boldmath$1$\unboldmath} \cdot \delta^{(3)}(
  \mbox{\boldmath$x$\unboldmath}-\mbox{\boldmath$x^\prime$\unboldmath})\;.
\end{equation}
The non-zero components of ${\cal M}^2(\phi,T)$ and 
$\mbox{\boldmath$\zeta$\unboldmath}$ are given by
\begin{eqnarray}
  {\mathcal{M}}_{11}^2&=&{\mathcal{M}}_{22}^2={\mathcal{M}}_{33}^2=
  {\mathcal{M}}_{66}^2=m_a^2(\phi)\;, \quad
  {\mathcal{M}}_{44}^2=m_\varphi^2(\phi,T)\;, \quad
  {\mathcal{M}}_{55}^2=m_h^2(\phi,T)\;,\nonumber\\
  \zeta_{14}^1&=&\zeta_{41}^1=1\;, \quad \zeta_{24}^2=\zeta_{42}^2=1\;,
  \quad \zeta_{34}^3=\zeta_{43}^3=1\;.
\end{eqnarray}
For the partial wave expansion of the
solution $\Psi(\mbox{\boldmath$x$\unboldmath})$ of the corresponding 
homogenous problem 
we use the following ansatz:
\begin{eqnarray}
\Psi(\mbox{\boldmath$x$\unboldmath})&=&\sum_{l=0}^{+\infty} 
\psi_l(\mbox{\boldmath$x$\unboldmath})\\
\label{psil}
\psi_l(\mbox{\boldmath$x$\unboldmath})&=&\sum_{m=-l}^{+l}
\left(  
\begin{array}{llllll}
\varphi_1^1(\mbox{\boldmath$\hat x$\unboldmath}) & 
\varphi_1^2(\mbox{\boldmath$\hat x$\unboldmath}) & 
\varphi_1^3(\mbox{\boldmath$\hat x$\unboldmath}) & 0 & 0 & 0\\
\varphi_2^1(\mbox{\boldmath$\hat x$\unboldmath}) & 
\varphi_2^2(\mbox{\boldmath$\hat x$\unboldmath}) & 
\varphi_2^3(\mbox{\boldmath$\hat x$\unboldmath}) & 0 & 0 & 0\\
\varphi_3^1(\mbox{\boldmath$\hat x$\unboldmath}) & 
\varphi_3^2(\mbox{\boldmath$\hat x$\unboldmath}) & 
\varphi_3^3(\mbox{\boldmath$\hat x$\unboldmath}) & 0 & 0 & 0\\
0 & 0 & 0 & \varphi_4^4(\mbox{\boldmath$\hat x$\unboldmath}) & 0 & 0\\
0 & 0 & 0 & 0 & \varphi_5^5(\mbox{\boldmath$\hat x$\unboldmath}) & 0\\
0 & 0 & 0 & 0 & 0 & \varphi_6^6(\mbox{\boldmath$\hat x$\unboldmath})\\
\end{array}
\right)
\left( 
\begin{array}{l}
f_1(r)\\f_2(r)\\f_3(r)\\f_4(r)\\f_5(r)\\f_6(r)\\
\end{array}
\right)\\
\varphi_j^1(\mbox{\boldmath$\hat x$\unboldmath}) &:=& 
\left[\mbox{\boldmath$Y$\unboldmath}_{l\  l+1\ m}
(\mbox{\boldmath$\hat x$\unboldmath})\right]_j\,,
\quad \varphi_j^2(\mbox{\boldmath$\hat x$\unboldmath}) := 
\left[\mbox{\boldmath$Y$\unboldmath}_{l\  l-1\ m}
(\mbox{\boldmath$\hat x$\unboldmath})\right]_j\,,
\quad \varphi_j^3(\mbox{\boldmath$\hat x$\unboldmath}) := 
\left[\mbox{\boldmath$Y$\unboldmath}_{l\  l\ m}
(\mbox{\boldmath$\hat x$\unboldmath})\right]_j\,,\\
\varphi_4^4(\mbox{\boldmath$\hat x$\unboldmath})&:=& 
\varphi_5^5(\mbox{\boldmath$\hat x$\unboldmath}) = 
\varphi_6^6(\mbox{\boldmath$\hat x$\unboldmath})=
Y_{lm}(\mbox{\boldmath$\hat x$\unboldmath})\,.
\end{eqnarray}
Here $\mbox{\boldmath$\hat x$\unboldmath}:=\mbox{\boldmath$x$\unboldmath}/r$ 
means the unit vector in $\mbox{\boldmath$x$\unboldmath}$--direction.
To shorten the notation the $l$-- and $m$--dependence
of the functions $\varphi_i^j$ and the $l$--dependence of the functions
$f_i$ is not shown explicitly in the following.
The vector spherical harmonics $\mbox{\boldmath$Y$\unboldmath}$ 
(see e.~g.\ \cite{Edmonds:1964})
are related to the common surface spherical harmonics:
\begin{eqnarray}
\mbox{\boldmath$Y$\unboldmath}_{l\ l+1\ m}&=& \frac{1}{c_1}\frac{1}{2l+1} 
\Bigl( -(l+1) \mbox{\boldmath$\hat x$\unboldmath} Y_{lm}
+r \mbox{\boldmath$\nabla$\unboldmath} Y_{lm}\Bigr)\;,\\
\mbox{\boldmath$Y$\unboldmath}_{l\ l-1\ m}&=& \frac{1}{c_0}\frac{1}{2l+1} 
\Bigl( l \mbox{\boldmath$\hat x$\unboldmath} Y_{lm}
+ r \mbox{\boldmath$\nabla$\unboldmath} Y_{lm} \Bigr)\;,\\
\mbox{\boldmath$Y$\unboldmath}_{l\ l\ m}&=& \frac{1}{\sqrt{l(l+1)}}{\bf I\!L} 
Y_{lm\;,}\;,\\
c_0&:=& \sqrt{\frac{l}{2l+1}}\;, \qquad c_1:=\sqrt{\frac{l+1}{2l+1}}\;.\\
\end{eqnarray}
For the partial wave $l=0$ one has to consider the following peculiarity:
neither the function 
$\mbox{\boldmath$Y$\unboldmath}_{0\ -1 0}(\mbox{\boldmath$\hat x
$\unboldmath})$ nor $\mbox{\boldmath$Y$\unboldmath}_{0\ 0\ 0}
(\mbox{\boldmath$\hat x$\unboldmath})$
exists, because the normalizing prefactor diverges. Scaling
$f_2(r)$ with $c_0$ and $f_3(r)$ with $\sqrt{l(l+1)}$, the products
$\mbox{\boldmath$Y$\unboldmath}_{l\ l-1\ m} f_2(r)$ and 
$\mbox{\boldmath$Y$\unboldmath}_{l\ l\ m} f_3(r)$ are well defined
also at $l=0$; they yield zero. I.~e., at $l=0$
the components $f_2(r)$ and $f_3(r)$ vanish identically.
Hence, the coupled $(4\times 4)$--system
reduces to a coupled $(2\times 2)$--system.

In the following the index $l$ of the functions $\psi$ is 
suppressed. Using the index notation the ansatz (\ref{psil}) 
can be formulated as
\begin{equation}
\psi_n=\varphi_n^i f_i\,,
\end{equation}
and with
\begin{equation}
{\cal D}_{nk} := \left( \frac{\partial^2}{\partial r^2} 
+ \frac{2}{r} \frac{\partial}{\partial r} - \frac{1}{r^2}{\bf I\!L}^2
\right)\delta_{nk} - {\cal M}_{nk}^2
\end{equation}
the homogeneous equation for $\Psi$ takes the form:
\begin{equation}
{\cal D}_{nk} \psi_k = g\phi^\prime \zeta_{nk}^i\frac{x_i}{r}\psi_k\;.
\end{equation}
Using
\begin{equation}
{\bf I\!L}^2 \varphi_n^i = l_i(l_i+1) \varphi_n^i \qquad \mbox{with} \qquad
l_i=\left(l+1 , l-1 , l, l, l, l \right)_i\,,
\end{equation}
we get
\begin{equation}
\label{equ:homogeneous}
{\cal D}_{nk} \psi_k = \varphi_k^i\left[\left( \frac{\partial^2}{\partial r^2} 
+ \frac{2}{r} \frac{\partial}{\partial r} - \frac{l_i(l_i+1)}{r^2}
\right)\delta_{nk} - {\cal M}_{nk}^2 \right] f_i = 
g\phi^\prime \zeta_{nk}^i\frac{x_i}{r}\psi_k\,.
\end{equation}
Let $j$ be an index that takes values from 1 to 3.
Then we have
\begin{equation}
\label{rhsdglpsi}
  {\cal D}_{nj} \psi_j + {\cal D}_{n4} \psi_4
  + {\cal D}_{n5} \psi_5+ {\cal D}_{n6} \psi_6 =
  g \phi^\prime \delta_{n4} \frac{x_j}{r} \psi_j + g \phi^\prime
  \delta_n^j\frac{x_j}{r} \psi_4\,.
\end{equation}
With the useful relations \cite[S. 103f.]{Edmonds:1964}
\begin{eqnarray}
  \frac{x_j}{r}\varphi_4^4 &=& - c_1 \varphi_j^1 + c_0\varphi_j^2\;,\\
  \frac{x_j}{r}\psi_j &=& \frac{x_j}{r}\varphi_j^i f_i = 
  \mbox{\boldmath$\hat x$\unboldmath} \left( f_1 
\mbox{\boldmath$Y$\unboldmath}_{l\ l+1\ m}+
    f_2 \mbox{\boldmath$Y$\unboldmath}_{l\ l-1\ m} +f_3 
\mbox{\boldmath$Y$\unboldmath}_{l\ l\ m}\right)\nonumber\\
  &=& -f_1c_1\varphi_4^4 + f_2c_0\varphi_4^4\,,
\end{eqnarray}
the right-hand side of (\ref{rhsdglpsi}) can be written as:
\begin{equation}
\label{rhsdglpsi2}
g\phi^\prime \delta_{n4} \left( -f_1 c_1 \varphi_4^4 + f_2 c_0 \varphi_4^4
\right) + g\phi^\prime \delta_n^j\left( -c_1 \varphi_j^1 + c_0\varphi_j^2
\right) f_4\;.
\end{equation}
Therefore, as the vector spherical harmonics are linear independent,
(\ref{equ:homogeneous}) only can be satisfied for $f_k$ being a solution of
\begin{eqnarray}
\label{equakr}
&&\left[\left(\frac{\partial^2}{\partial r^2}+\frac{2}{r}
\frac{\partial}{\partial r} - \frac{l_n(l_n+1)}{r^2} \right) \delta_{nk}
-{\mathcal{V}}_{nk}(\phi)\right] f_k(r)=0\,,\\
\mbox{with }&~& {\mathcal{V}}(\phi)=
\left(
\begin{array}{llllll}
m_a^2(\phi) & 0 & 0 & -c_1g\phi^\prime & 0 & 0\\
0 & m_a^2(\phi) & 0 & c_0g\phi^\prime & 0 & 0\\
0 & 0 & m_a^2(\phi) & 0 & 0 & 0\\
-c_1g\phi^\prime & c_0g\phi^\prime & 0 & m_\varphi^2(\phi,T) & 0 & 0\\
0 & 0 & 0 & 0 & m_h^2(\phi,T) & 0\\
0 & 0 & 0 & 0 & 0 & m_a^2(\phi)\\
\end{array}
\right)\;.
\end{eqnarray}
These considerations motivate the following ansatz for the
Green's function $G(\mbox{\boldmath$x$\unboldmath},
\mbox{\boldmath$x^\prime$\unboldmath})$:
\begin{equation}
G_{km}(\mbox{\boldmath$x$\unboldmath},\mbox{\boldmath$x^\prime$\unboldmath}) =
\sum_{l,m} g_{rs}(r,r^\prime) \varphi_k^r(\mbox{\boldmath$\hat x$\unboldmath})
\varphi_m^{s\ast}(\mbox{\boldmath$\hat x^\prime$\unboldmath})\,,
\end{equation}
where the radial Green's functions -- their $l$--dependence is not 
displayed explicitly -- satisfy
\begin{equation}
  \label{radial_greenfkt}
  \left[\left(\frac{\partial^2}{\partial r^2}+\frac{2}{r}
      \frac{\partial}{\partial r} - \frac{l_n(l_n+1)}{r^2} \right) \delta_{nk}
    -{\mathcal{V}}_{nk}(\phi)\right] g_{km}(r,r^\prime)=
  -\frac{1}{r^2}\delta(r-r^\prime) \delta_{nm}\;.
\end{equation}
In (\ref{funcder_genform}) one needs to know the following expressions:
\begin{eqnarray}
  {\rm tr} \left[ c \; \frac{\partial {\cal M}^2}{\partial \phi}
    G\right] &=& \sum_{l=0}^{+\infty} \frac{2l+1}{4\pi} 
  \left \{
    \frac{3g^2}{4}\phi\left( g_{11} + g_{22} + g_{33}\right) 
    + \frac{3}{2}\left(\frac{g^2}{2}\phi + 2\lambda_T\phi-3ET\right)g_{44}
  \right.\nonumber\\
  && 
  \left.
    + 3 (\lambda_T \phi -ET ) g_{55}
    -\frac{3g^2}{4}\phi g_{66}
  \right \}\;,\\
  {\rm tr} \left[ c \; \mbox{\boldmath$\zeta$\unboldmath} 
\mbox{\boldmath$\nabla$\unboldmath} G \right] &=& 
  \frac{3}{2} \sum_{l,m}
  \zeta_{nm}^r \nabla_r g_{ab}\varphi_m^a \varphi_n^{b\ast}
  = \frac{3}{2}
  \sum_{l,m} \left( \nabla_i g_{4j} \varphi_4^4 \varphi_i^{j\ast}
    +\nabla_i g_{j4} \varphi_i^j \varphi_4^{4\ast}\right)\nonumber\\
  &=& \frac{3}{2} \sum_l \frac{2l+1}{2\pi} \left\{ -c_1 \left( g_{14}^\prime
      +\frac{2}{r}g_{14}\right) + c_0 
    \left( g_{24}^\prime+\frac{2}{r}g_{24}\right)
  \right\}\;.
\end{eqnarray}
Obviously the components $g_{33}$ and $g_{66}$ satisfy the same
differential equation (see (\ref{radial_greenfkt})).
Hence, they can be combined. Due to the different sign of their
factors of degeneracy they cancel exactly, except for the s--wave.
As mentioned above at $l=0$ the amplitudes $f_2(r)$ and $f_3(r)$
vanish and therefore $g_{33}\equiv 0$ in this particular partial wave.
Hence, the $(6 \times 6)$-- reduces to a 
$(5 \times 5)$--system, where the fifth channel only contributes at
$l=0$. The final result is given in (\ref{funcder_result}).
\end{appendix}
%
%

%
%
\section*{Tables}
\begin{center}
\begin{tabular}{|r|r|r|r|r|}
\hline
&&&&\\[-0.3cm]
$m_H$ & $T_0^{\rm ht}$ & $T_c^{\rm ht}$ & $T_0^{\rm R}$ & $T_c^{\rm R}$\\
\hline
30 & 67.572 & 68.238 & 67.607 & 68.070\\
40 & 74.811 & 75.464 & 74.916 & 75.433\\
\hline
\end{tabular}
\end{center}
\begin{quotation}
  \noindent Table~1:
  The values of the roll-over temperature $T_0$ and the critical 
  temperature $T_c$ obtained from the high temperature potential $V_{\rm ht}$
  and the approximation of the one-loop resummed effective potential
  $V_{\rm R}$. The top-quark mass is chosen as $m_t=170$GeV. The Higgs boson
  mass $m_H$ and the temperatures are given in GeV.
\end{quotation}
\nopagebreak

\section*{Figure captions}
\noindent {\bf Figure 1:}
The expansion of the one-loop correction $S_{i,\rm R}$ to the
effective action. The lines represent the propagators and the dots the
vertex factors given by ${\cal U}_i(\phi,T)-{\cal U}_i(0,T)$.\medskip

\noindent {\bf Figure 2:}
Comparison of the logarithm of the nucleation rate $\gamma$
(see (\ref{nuclrate})),
i.~e. including the one-loop corrections and prefactors,
versus the temperature $T$. 
$\gamma$ has dimension 
$\mbox{GeV}^4$ and is calculated in units of $(g\widetilde v(T))^4$. 
The squares represent the results
we obtain using the profile
$\phi_{\rm cl}$, which is a solution of (\ref{classprofequ}). They
are connected by straight dashed lines.
The dots are the
results we obtain using the the self-consistent solution of the
full one-loop field equation (\ref{fieldequation3}). The solid
line is the function $F(T)$ as defined in (\ref{func:F_t}), 
with $a=3.48\cdot 10^{-2}$
and $b=1.350$. The parameter set used here is
$m_H=30$GeV and $m_t$=170GeV.\medskip

\noindent {\bf Figure 3:}
The same as in figure~2 for the parameter set
$m_H=40$GeV and $m_t$=170GeV. Here the solid line is the function
$F(T)$ as defined in (\ref{func:F_t}), with $a=2.38\cdot10^{-2}$ 
and $b=1.362$.\medskip

\noindent {\bf Figure 4:}
Comparison of the logarithm of the nucleation rate after
mapping of the temperature intervals $[T_0^{\rm R},T_c^{\rm R}]$ and 
$[T_0^{\rm ht},T_c^{\rm ht}]$ to $[0,1]$.
For the parameter set $m_H=30$GeV ($m_H=40$GeV) and $m_t=170$GeV
the squares (triangles) are the results obtained
when using the profile $\phi_{\rm cl}$ and
the dots (diamonds) are due the self-consistent profile.\medskip

\noindent {\bf Figure 5:}
The logarithm of the fraction $\rho_a(t)$
of space that has been converted to the asymmetric phase in
dependence of the time $t$. The three lines are due to three
different values of the bubble wall velocity $v$. The solid line
is obtained for $v=c$, the short-dashed line for $v=c/10$ and
the long-dashed one for $v=c/1000$. The 
parameter set used here is $m_H=30$GeV and $m_t=170$GeV.\medskip

\noindent {\bf Figure 6:}
The total number of bubbles per unit volume $N(t)$
in dependence of the time $t$. The three lines are due to three
different values of the bubble wall velocity $v$. The description
is the same as in figure~5.\medskip

\noindent {\bf Figure 7:}
The average bubble radius $R(t)$
in dependence of the time $t$. The three lines are due to three
different values of the bubble wall velocity $v$. The description
is the same as in figure~5.
\clearpage

\centerline{\Large Figure 1}
\begin{center}
  \setlength{\unitlength}{1.2cm}
  \begin{picture}(12,2)
    \thicklines
    \put(0.3,0.7){\Large $S_{i,\rm R}= \quad + c_i$}
    \multiput(3.5,0.9)(3,0){3}{\circle{1.2}}
    \multiput(3.5,0.3)(3,0){3}{\circle*{0.2}}
    \put(6.5,1.5){\circle*{0.2}}
    \put(8.98,1.2){\circle*{0.2}}
    \put(10.02,1.2){\circle*{0.2}}
    \put(4.6,0.7){\Large $-\quad \frac{c_i}{2}$}
    \put(7.6,0.7){\Large $+ \quad \frac{c_i}{3}$}
    \put(10.6,0.7){\Large $+ \dots$}
  \end{picture}
\end{center}

\begin{center}
\begin{picture}(12,9)
  \put(0,-0.5){\epsfxsize=12cm\epsfbox{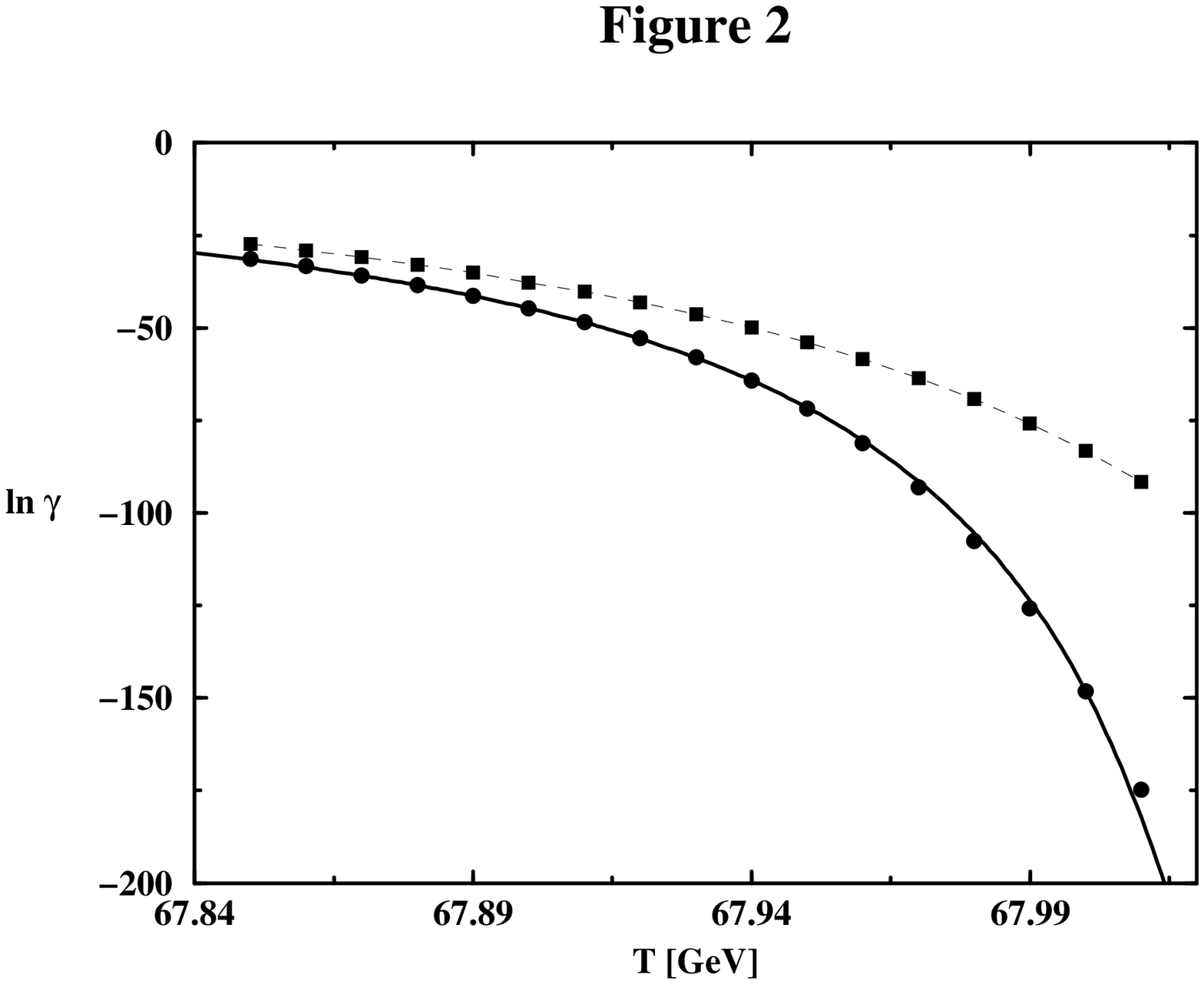}}
\end{picture}
\end{center}

\begin{center}
\begin{picture}(12,9)
  \put(0,-0.5){\epsfxsize=12cm\epsfbox{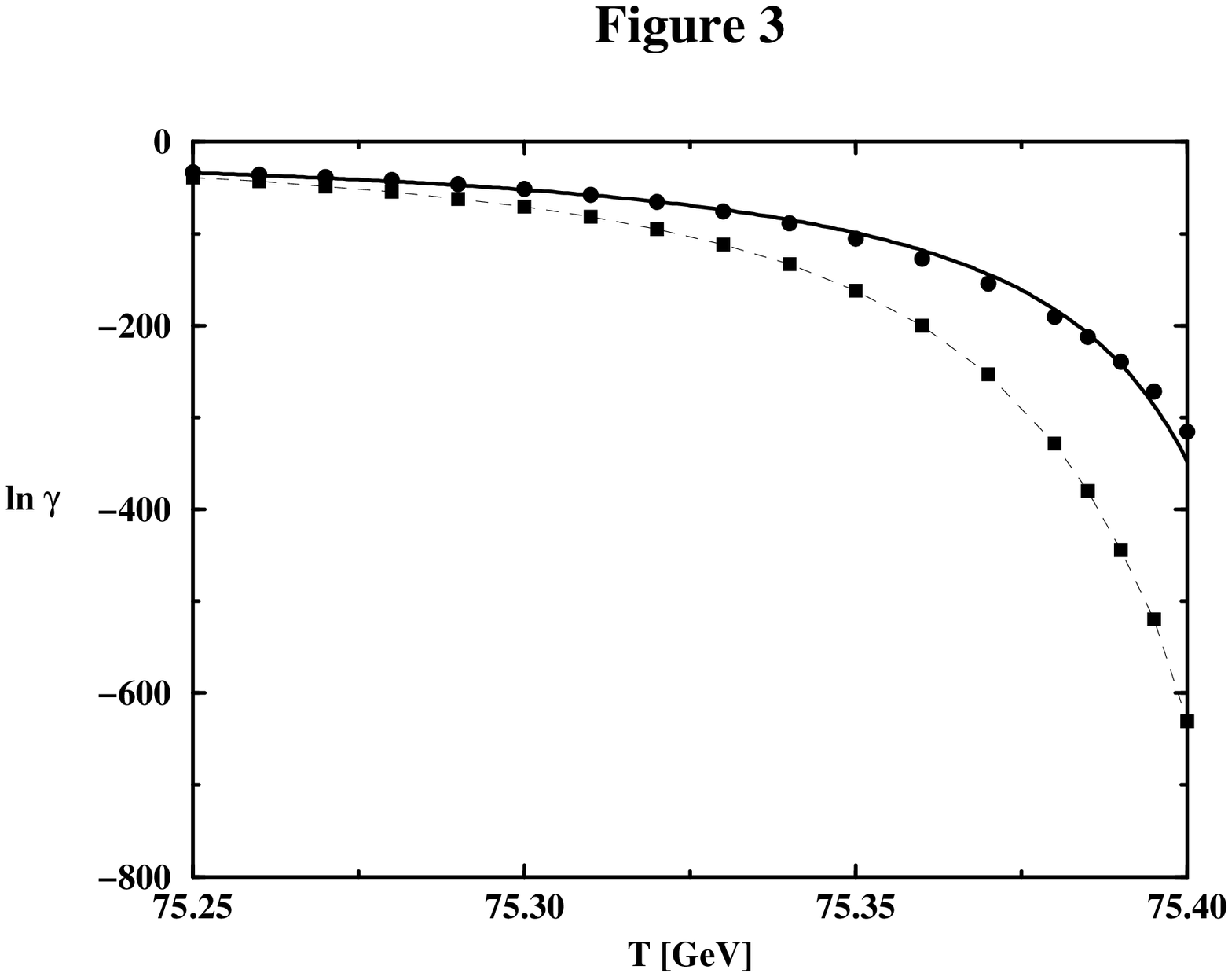}}
\end{picture}
\end{center}
\clearpage

\begin{center}
\begin{picture}(12,9)
  \put(0,-0.5){\epsfxsize=12cm\epsfbox{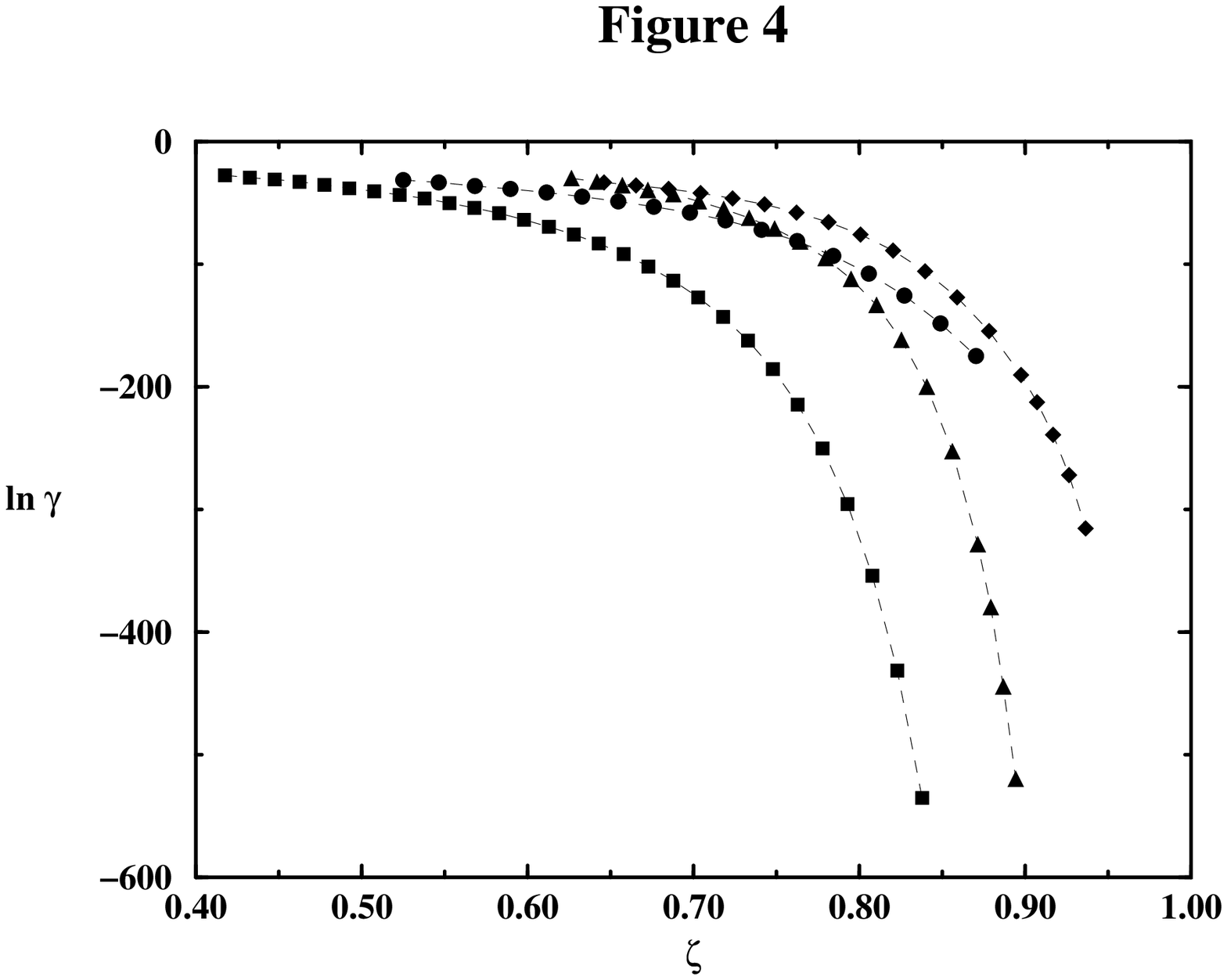}}
\end{picture}
\end{center}
\vspace{3cm}

\begin{center}
\begin{picture}(12,9)
  \put(0,-0.5){\epsfxsize=12cm\epsfbox{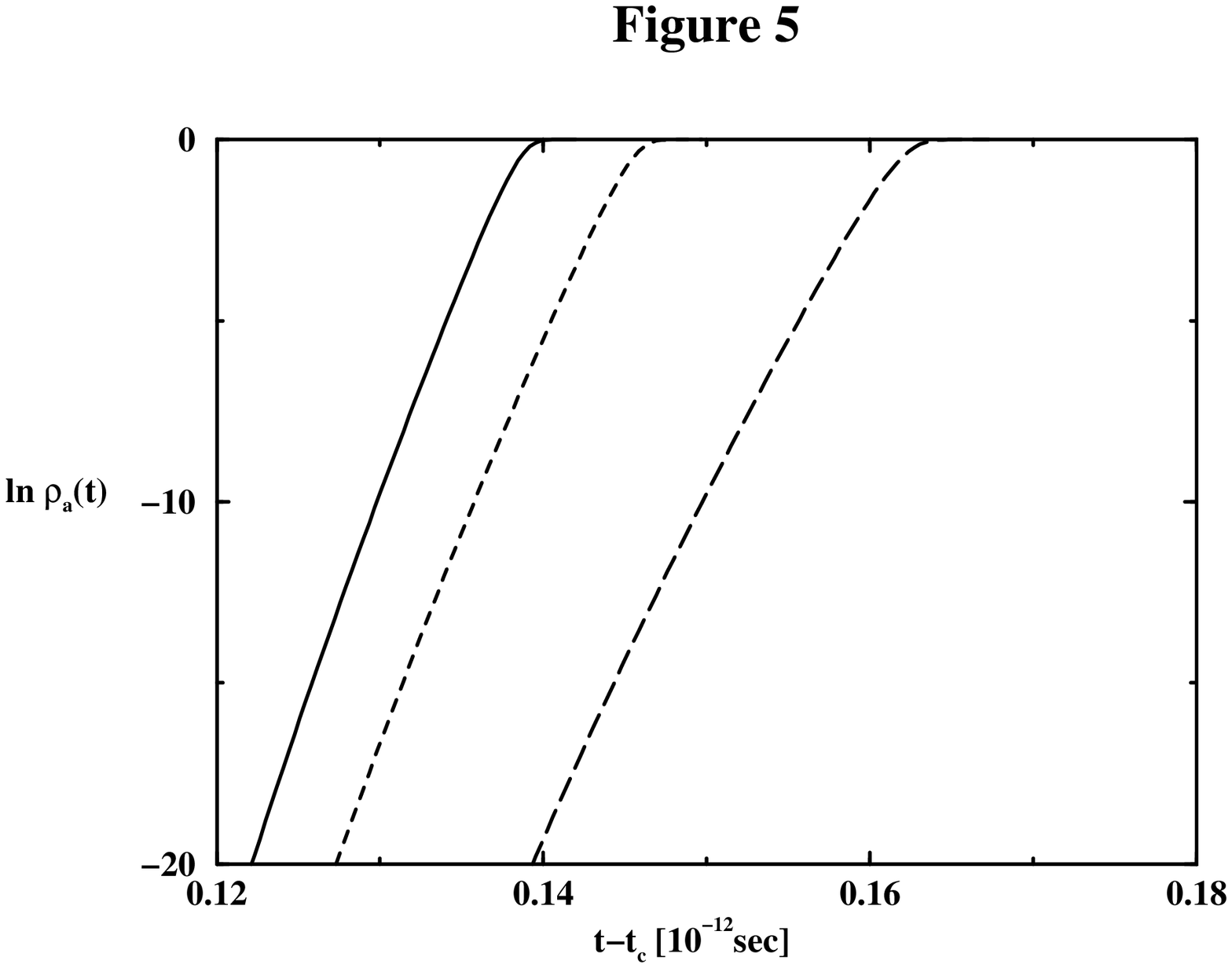}}
\end{picture}
\end{center}
\clearpage

\begin{center}
\begin{picture}(12,9)
  \put(0,-0.5){\epsfxsize=12cm\epsfbox{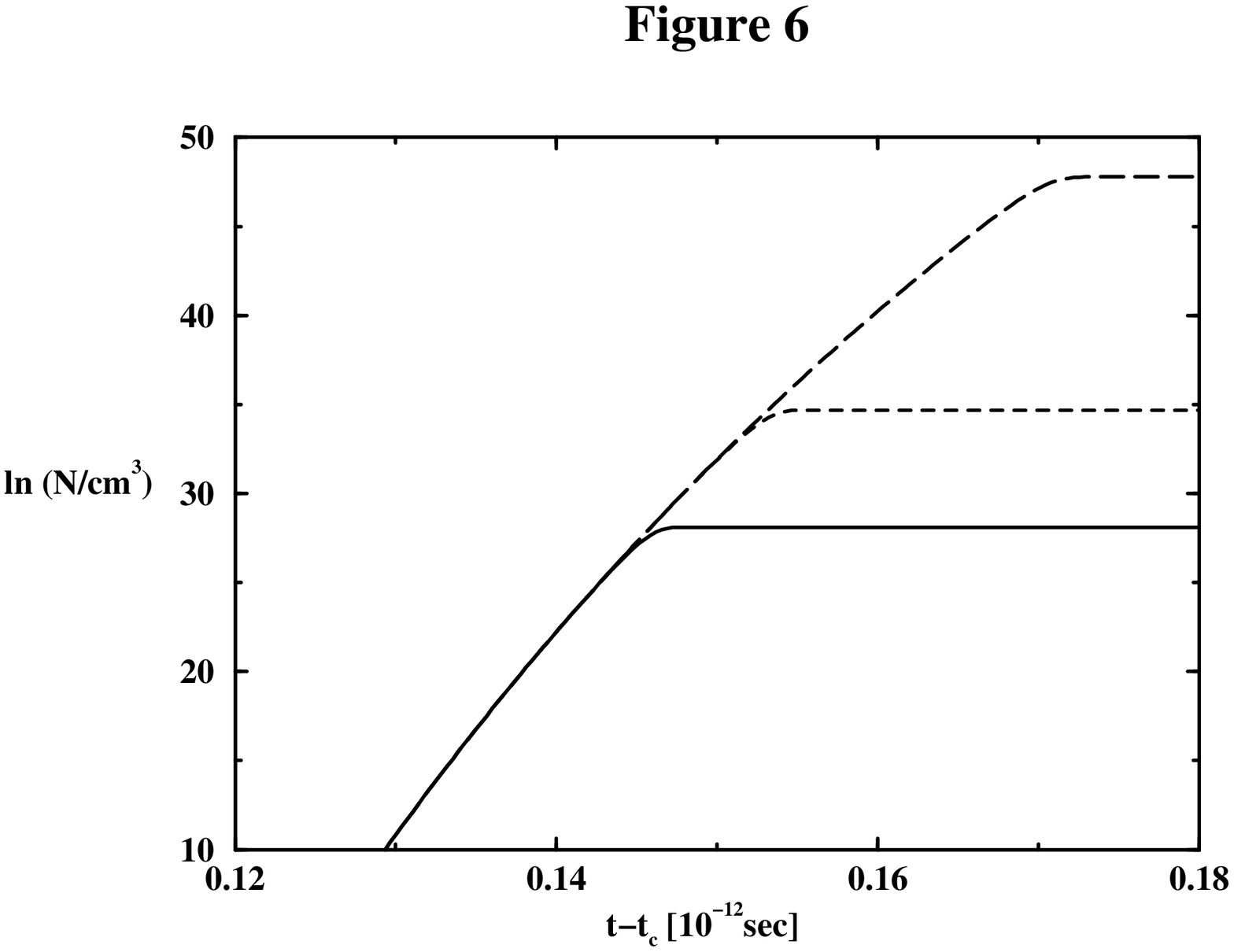}}
\end{picture}
\end{center}
\vspace{3cm}

\begin{center}
\begin{picture}(12,9)
  \put(0,-0.5){\epsfxsize=12cm\epsfbox{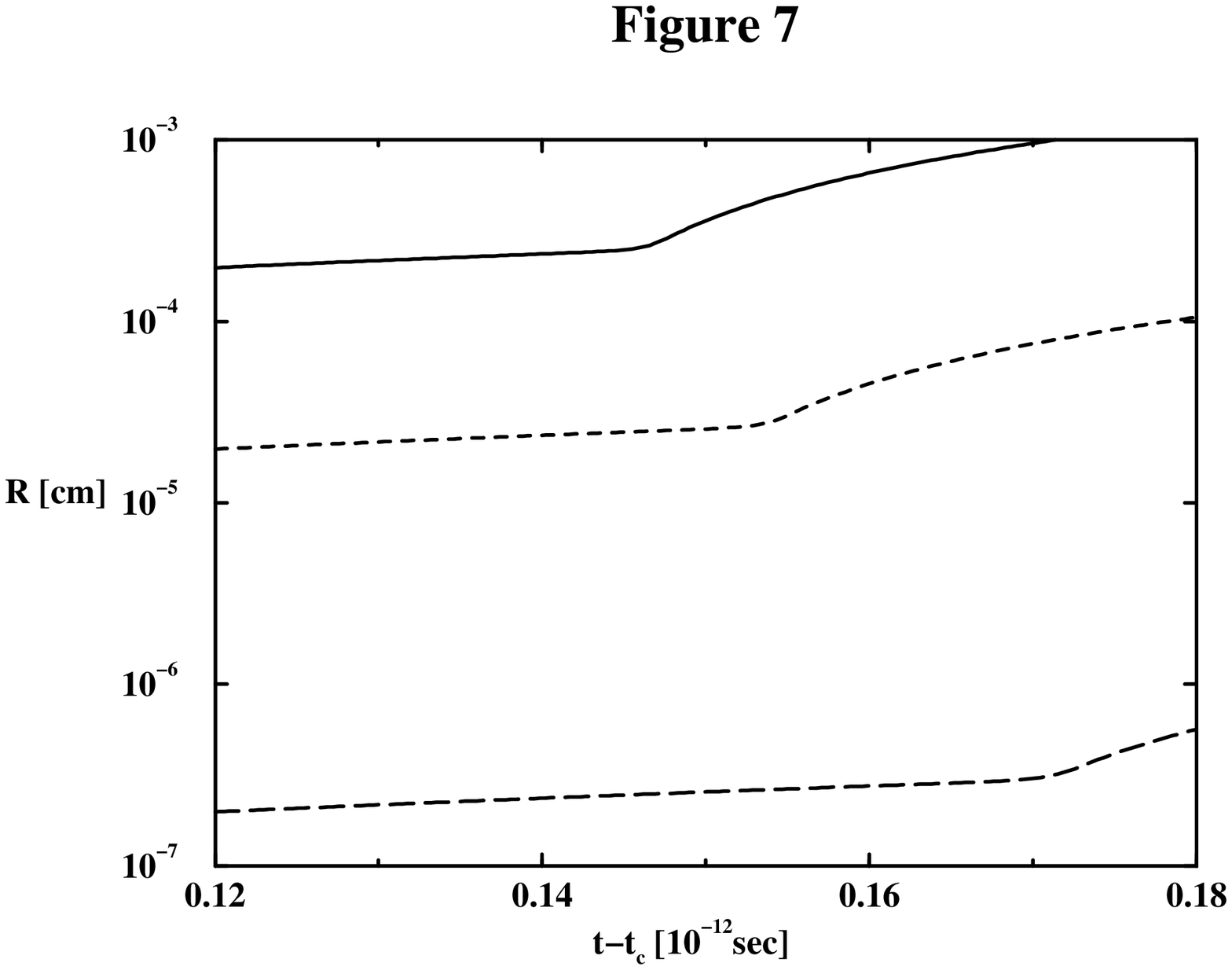}}
\end{picture}
\end{center}
\end{document}